\documentstyle[preprint,aps]{revtex}
\begin{document}
\draft
\preprint{ WC -- \today}
\title{Variational quantum Monte Carlo study of \\
two-dimensional Wigner crystals:\\
exchange, correlation, and magnetic field effects}
\author{Xuejun Zhu$^{1,2}$ and Steven G. Louie$^2$}
\address{
$^1$Department of Physics \& Astronomy,
Rutgers University, Piscataway, NJ 08855 \\
$^2$Department of Physics, University of California, and, \\
Materials Sciences Division, Lawrence Berkeley Laboratory, Berkeley,
                 CA 94720 }
\date{\today}
\maketitle
\begin{abstract}

The two-dimensional Wigner crystals are studied with the
variational quantum Monte Carlo method.
The close relationship between the ground-state
wavefunction and the collective excitations in the system is
illustrated, and used to guide the construction of the ground-state
wavefunction of the strongly correlated solid.
Exchange, correlation, and magnetic field effects all give rise to
distinct physical phenomena.
In the absence of any external magnetic field,
interesting spin-orderings are observed in the ground-state of
the electron crystal in various two-dimensional lattices. In particular,
two-dimensional bipartite lattices are shown not to lead necessarily
to an antiferromagnetic ground-state.
In the quantum Hall effect regime,
a strong magnetic field introduces new energy and length scales.
The magnetic field quenches the kinetic energy and poses
constraints on how the electrons may correlate with each other.
Care is taken to
ensure the appropriate translational properties of
the wavefunction when the system is in a uniform
magnetic field. We have examined the exchange,
intra-Landau-level correlation as well as Landau-level-mixing effects
with various variational wavefunctions.
We also determine their dependences
on the experimental parameters such as the carrier effective mass
at a modulation-doped semiconductor heterojunction. Our
results, when combined with some
recent calculations for the energy of the
fractional quantum Hall liquid including Landau-level-mixing,
show quantitatively that in going from $n$-doping to $p$-doping
in $GaAS/AlGaAS$ heterojunction systems,
the crossover filling factor
from the fractional quantum Hall liquid to the Wigner
crystal changes from filling factor $\nu \sim 1/5$ to $\nu \sim 1/3$.
This lends strong support to the claim that the observed
reentrant insulating phases around $\nu = 1/5$ for $n$-doped
and around $\nu = 1/3$ for $p$-doped high-mobility
samples are primarily caused by electron-electron interaction effects.
We discuss the possible implications of our theoretical
results for some recent experiments carried out
in the quantum Hall regime in search of the electron solid.

\end{abstract}

\newpage
\section{Introduction}

In this paper, we present a comprehensive study of the
various aspects, mostly related to the
ground-state energies, of the properties
of the two-dimensional (2D) electron Wigner crystal (WC) \cite{Wigner}.
Numerical calculations are carried out with the variational quantum
Monte Carlo (VMC) method.
This approach, pioneered by McMillan for
$^4He$ systems \cite{VMC}, has been used extensively to study
many fermion systems
\cite{CeperleySci,Fahy,Ceperley,Zhu,Price93}.
The purpose of the present work is two-fold. One is to examine
how to construct a good variational
ground-state wavefunction for the electron solid
by exploring the intimate relation between the
ground-state wavefunction and the collective
excitation properties.
This construction is followed both in the presence
and in the absence of a
strong perpendicular magnetic
field. The second goal is to study the fractional quantum Hall
liquid-Wigner crystal transition by calculating accurately
the energies of the Wigner crystal. In doing
so, we also obtain a quantitative understanding of the
sizes of various interaction effects such as exchange,
intra-Landau-level correlation, and inter-Landau-level correlation.
We make contact with some recent experiments
by comparing
the present WC energy to the quantum Hall liquid energy
\cite{Zhu,Price93,Rappe,Ortiz}.
By using different
wavefunctions in the Monte Carlo calculations, we study various aspects
of the physical properties of the electron solid. We find that
exchange, correlation, and magnetic field effects all give rise to
some interesting physical phenomena.
Some of our results have been reported in a
short paper \cite{Zhu}.

In addition to the intrinsic theoretical
interest of the properties of a Wigner crystal, our work
has been directly
motivated by the recent experimental activities
looking for signatures of the electron solid in the
integer quantum Hall effect (IQHE) and the fractional
quantum Hall effect (FQHE) regime
\cite{FQHE0,TheBook,Halperin83,Jiang1,I,Andrei,Willett,II,III,MagnetoOp,GapCollapse,Jiang,hole,WCSi}.
Exchange-correlation effects in these 2D electron systems in
a strong magnetic field are
different in essential ways from those in zero field.
Most importantly, correlation-induced fluctuations are
allowed, at fractional filling factors,
to occur within the same Landau-level
at no cost to the kinetic energy.
These nearly ideal 2D systems
exhibit a very rich variety of quantum phases
and phase transitions
\cite{Halperin83,CSGL,Platzman93}.
Furthermore, the phases and the phase transitions can be controlled
and studied experimentally,
by changing the carrier density, the carrier effective mass, the
field strength, and in some cases, the number of 2D
layers involved.
At present, we are only beginning to assess the quantitative
aspects of this interesting phase diagram
\cite{Zhu,Price93,Rappe,Ortiz,Platzman93}.

In the rest of this Introduction, we make some general remarks on
the problem of Wigner crystallization, discuss
some recent experimental work that stimulated our
investigation, and summarize our main
results.

\subsection{General Remarks}

It has long been expected theoretically that at $T = 0$, an
interacting electron system in a uniform positive
background (a jellium model),
will undergo a transformation from a liquid to a solid phase as its
density is lowered \cite{Wigner}.
The Hamiltonian of the model jellium system is  simply
(in atomic units and in the absence of external fields):
\begin{equation}
H = \sum_i {-\nabla_i^2 \over 2}
+\sum_{i \ne j} { 1\over 2r_{ij}},
\label{Hamil}
\end{equation}
where interaction with a neutralizaing background is implied.
The idea of the Wigner crystallization is quite
intuitive. The system is characterized by
the density parameter $r_s$ (defined in 2D
by its density $n$ measured in atomic units through $\pi r_s^2 = 1/n$).
Roughly speaking, the kinetic energy
of the system scales as $1/r_s^2$ and the Coulomb interaction energy
scales as $1/r_s$. In a normal metal, $r_s$ is on the order of 1 and
the kinetic energy is more important. The system is therefore characterized
by the Fermi liquid theory with a well defined Fermi surface. However, if one
were able to make $r_s$ arbitrarily large, there ought to be a crossover to
a regime where the interaction becomes dominant.
The
resultant state is then one in which electrons are localized in a
close-packed-lattice so that the average distance between them
is maximized. In the total absence of the
kinetic energy, the ground-state configuration of the
electrons will correspond to
the global minimum of the interaction potential. Properties of such
an electron solid in general will not be obtainable from perturbative
considerations around the liquid state:
before and after the solidification, both the collective excitations and
the single-particle excitations are qualitatively different.
For example, in the solid phase, there will be a gap of
the order of $e^2/r_s$ to
single-particle excitations, and there will be resistance
to shear. Neither occurs in the liquid phase.

Despite its theoretical certainty, direct experimental
observation of the Wigner crystallization has been difficult.
Only partial realization of Wigner's proposal
has been achieved in a system of two-dimensional
electrons trapped
on the surface of liquid He \cite{GrimesAdams}.
In this experiment, the areal
densities of the 2D electrons,
ranging from $10^{5}$ to $10^{9}~cm^{-2}$,
are so low that the Fermi energy is always more
than an order of magnitude smaller than the
temperature at which these experiments are carried out.
Therefore
the system is essentially a classical one-component plasma.
Nonetheless, when the interaction energy dominates over the
kinetic energy, which is simply $\sim kT$ in this
classical regime, one could
still observe the Wigner crystallization
\cite{GrimesAdams}.
In a three-dimensional system ($n$-doped $HgCdTe$),
by measuring the magneto-resistance and the Hall
resistance, it has been suggested that magnetically
induced three-dimensional Wigner crystallization may have been realized
\cite{Rosenbaum}.

By far the most intense experimental work
in the pursuit of Wigner crystals has been carried out in
quantum Hall devices
\cite{FQHE0,TheBook}. This regime is also more
interesting since the competing liquid phase is a strongly
correlated quantum Hall liquid which exhibits unusual
transport properties \cite{FQHE0,Laughlin}.
Modulation doped $GaAs/AlGaAs$ heterojunctions
and silicon inversion layer devices provide an
almost ideal experimental
realization of a two-dimensional jellium system.
Compared with electrons on helium,
the lower carrier effective mass,
higher density, and lower temperatures
place the system in the quantum
regime.
Mobilities of $GaAs/AlGaAs$ samples can be
as high as $10^7$ $cm^2/(V s)$, corresponding to an effective
mean-free-path
almost of macroscopic length ($\sim$ 0.05 $mm$).
For hole-doped samples, mobilities are somewhat
lower, $\sim 10^5 - 10^6$ $cm^2/(V s)$ \cite{hole}.
In silicon MOSFETs, mobilities are still lower, around
$10^4-10^5$ $cm^2/(V s)$, making them less ideal \cite{WCSi}.
It has long been suggested that the application
of a strong perpendicular magnetic field, which
quenches the kinetic energy and confines
the electrons to the size of the
magnetic length $l_B = \sqrt{\hbar c/eB}$,
will facilitate the electron crystallization \cite{Chaplik}.
Search along this direction culminated in the observation of
an unexpected
collective liquid state, {\it i.e.,\/}
the fractional quantum Hall liquid,
characterized by vanishing longitudinal resistance and
fractionally quantized Hall
resistance \cite{FQHE0,Laughlin}.

The phase diagram of a two-dimensional electron system in a strong
magnetic field is made much more intricate by the presence of
such quantum Hall states.
It is however
still expected that ultimately Wigner crystallization
would occur in a strong enough $B$-field, or a small enough Landau-level
filling factor $\nu = {{2\pi\hbar c n}\over eB}$.
In the last several years, many claims have
been made, suggesting the possible observation of the Wigner
crystal in both the modulation
doped $GaAs/AlGaAs$ and the silicon
inversion layer systems
\cite{Jiang1,I,Andrei,Willett,II,III,MagnetoOp,Jiang,hole,WCSi}.
We defer a more detailed discussion of the current
experimental situation to the next subsection.

Theoretical estimates of the crystallization density of a jellium
system free of external magnetic fields have varied widely in
the past \cite{CareMarch}.
The most reliable results are those given by the Green's function
Monte Carlo method
\cite{CeperleySci,Ceperley}.
The Monte Carlo studies have for the most part focused on
the body-centered-cubic lattice in 3D and the hexagonal lattice
in 2D \cite{CeperleySci,Ceperley,ZhuLouie,ZhuandLouie}.

In cases where a strong magnetic field is involved,
the competing liquid phase is the FQHE.
Quantitative estimates for the WC transition
filling factor in the quantum Hall regime
can be obtained by comparing the
energies of the Wigner crystal to those of the quantum Hall liquid.
Various approaches have been used for this purpose. On the liquid side,
exact diagonalization of small clusters with typically less than
ten particles has provided much insight into the nature of the
incompressible FQHE states
\cite{Haldane}. But extrapolations to the thermodynamic
limit for
quantities
like the ground-state energy
have so far proven inaccessible with this approach. Therefore
the most reliable energies for the FQHE states are obtained
variationally with Laughlin's trial wavefunction which is considered
very accurate \cite{Levesque}.
Price, Platzman, and He have recently reported
variational calculations of the FQHE liquid energies with a
Landau-level-mixing Jastrow factor on a sphere \cite{Price93}.
Variational Monte Carlo calculations
with planar (modified) periodic boundary
condition geometry have also been carried out \cite{Rappe}.
Both calculations give identical energies when the same
trial wavefunctions are used.

On the Wigner crystal
side, strictly variational approaches have been
largely limited
to the Hartree-Fock approximation which ignores
the crucial correlation effects \cite{WCHF,LLMHF}.
Perturbative phonon
treatments
beyond the harmonic level
for the electron solid
have been reported, but it remains unclear
how fast the higher-order phonon contributions converge
in the regime of density and magnetic field of
experimental interest \cite{Chui91}.
One exception that conbines the virtues of the two approaches
\cite{LamGirvin}
is the work by Lam and Girvin, where they optimized the
variational parameters in the trial wavefunction with a
truncated harmonic Hamiltonian, and then evaluated the
expectation value of the original Hamiltonian with such a
trial wavefunction. It captures most of the intra-Landau-level
correlation, but
does not treat the exchange effects, or the inter-Landau-level
correlations.
In addition, it has been
noted that all the odd-terms in the expansion of the
Hamiltonian in the phonon coordinates are not included in
the total energy due to the form of the harmonic trial wavefunction
\cite{ChuiOddTerms}.
The size of the third order term has been estimated
\cite{Chui91,ChuiOddTerms}.

If one compares the Wigner solid energy from Lam and Girvin in
Ref.~\cite{LamGirvin} with that
of the FQHE liquid by Levesque, {\it et al.}, in Ref.~\cite{Levesque},
the solid is favored for
$\nu \leq 1/6.5$. However, the
energy of the liquid at an arbitrary filling factor will be
higher than that from interpolating between the odd-denominator
filling factors \cite{Halperin}.
This gives rise to the possibility of a reentrant
WC-FQHE-WC transition as $\nu$ changes from $\nu > 1/5$ to
$\nu = 1/5$ to $\nu < 1/5$ as the magnetic field increases.
Such reentrant phase transitions around $\nu = 1/5$ have indeed
been observed by a variety of techniques and groups, and have
been mostly attributed to this mechanism
\cite{Jiang1,I,Andrei,Willett,II,III,MagnetoOp,Jiang,hole,WCSi}.

As we will discuss in the next subsection, some recent experiments
have now taken us to a regime where Landau-level-mixing
can not be realistically ignored \cite{hole}. In our work, we have
treated exchange, intra-Landau-level correlation, and
Landau-level-mixing all on equal footing \cite{Zhu}. It is hoped that
through our work one may develop a quantitative feeling for the
relative size of these effects under various experimental
conditions.

\subsection{Summary of the recent experiments}

Here, we
give a brief overview of some recent experimental activities
that are designed to detect the WC in
the FQHE regime
\cite{Jiang1,I,Andrei,Willett,II,III,MagnetoOp,Jiang,hole,WCSi}.
Since the experiments are still evolving
rapidly, our summary is necessarily incomplete and we apologize
for any unintentional omissions.

With the very first observation of the fractional
quantum Hall effect at $\nu = 1/3$, experimentalists
also encountered
an insulating phase that set in at a smaller
filling factor \cite{FQHE0}. As sample quality improves, this ``critical''
filling
factor has been pushed toward smaller values
continuously. It is therefore
clear that these early-observed insulating phases are due to
disorder-induced
localization, and not due to the interaction-induced Wigner
crystallization intrinsic to a disorder-free
2D electron gas.

In the last several years, however, evidence has emerged that
there are at least two insulating phases around the fractional
quantum Hall state $\nu = 1/5$
\cite{Jiang1,I,Andrei,II,III,MagnetoOp,Jiang}.
Furthermore,
with regards to the
electron doping concentration
and sample quality, these reentrant insulating phases are
much more robust than the earlier insulating phases; and they
become more pronounced when impurity effects are made weaker and/or
when interaction effects are made stronger.
In the best samples currently available (as judged
by the transport gap of the FQHE state at $\nu = 1/5$ and by
the sample mobility), the insulating phases at before and after
$\nu = 1/5$ still persist \cite{Jiang}. The one
at $\nu = 0.21$ even grows in strength with sample
quality, as seen from the size of the insulating gap deduced
from transport measurements.
This has led to the conclusion
that these insulating phases are not
due to disorder
\cite{Jiang}. Since disorder is not strong enough to
destroy the FQHE state at $\nu = 1/5$, it is unlikely to
localize all the electrons at $\nu \geq 1/5$. Recall that
the magneto-roton
gap is very small for
the $\nu = 1/5$ FQHE state, and thus it
is very susceptible to
disorder \cite{GirvinRoton}.

Various experimental techniques have been used to study the
insulating phases around $\nu = 1/5$. These include
the traditional magneto-transport
\cite{Jiang1}, radio-frequency absorption \cite{Andrei},
surface acoustic wave absorption
\cite{Willett}, nonlinear (AC and DC) transport \cite{I,II,III,Jiang},
noise generation \cite{I,III},
magneto-optics \cite{MagnetoOp},
{\it etc.} The list is not exhaustive and
is still growing.
While they have revealed many interesting
properties of the insulating phases, they have also brought about
some controversies.

The traditional magneto-transport establishes the existence of
the insulating phases \cite{Jiang1}.
Unfortunately, it does not tell us
directly what gives rise to the insulating behavior. The clear
stability of the $2/9$ and $1/5$ FQHE states suggests that
interaction is more important than disorder \cite{Jiang1}. Radio-frequency
absorption \cite{Andrei}, which attempts to measure the dispersion of the
lower-hybrid magneto-phonon, has been fit to the characteristic
$q^{3/2}$-dispersion, but later was fit to a $q^{1/2}$-mode. The
latter dispersion is more likely in the presence of disorder. However
it has been argued that the data can still be fit with a linear-in-$q$
dispersion \cite{Andrei}.
This uncertainty leaves doubt as to the reliability of
the interpretation of the experimental results. At least, the range
of $q$ that was covered by the experiment is not wide enough
to establish unequivocally the dispersion.
The surface acoustic wave
absorption data \cite{Willett},
while giving the important frequency dependence
of the collective mode, make clear that in the $q$-range studied,
the collective modes are very broad and unable to relate them
rigorously to the existence of a Wigner crystal.

Nonlinear transport measurements have provided yet another way
to study the solid \cite{I,II,III,Jiang}.
A collective sliding motion would be a
signature of a Wigner crystal with reasonable coherence,
similar to a charge-density wave system \cite{CDWRev}.
But at present results from various groups differ
in important details. For example, the sliding threshold
measured in two experiments using samples with
similar mobility differed by a factor of $\sim 500$.
The differential conductivity above the sliding threshold
became field-independent in some, but remained
field dependent in others.
A recent theoretical work has sought to
unify some of the experimental results where a detailed
comparison and analysis of the nonlinear transport experiments
can be found \cite{XJZPBLAJM}. The possible $AC-DC$
interference effects in the sliding state have also been
explored both experimentally \cite{III}
and theoretically \cite{XJZPBLAJM}. Clear Shapiro steps
or an inductive anomaly at a well-defined AC frequency
would constitute strong evidence for the crystalline
order in the insulating phases \cite{XJZPBLAJM}.

More direct spectroscopic tools have also
been brought to bear on the FQHE/WC problem.
Magneto-optical measurements have revealed several
interesting aspects of the system \cite{MagnetoOp,GapCollapse}.
First of all,
spectroscopic features, rather similar to those at
$\nu = 1/3, 2/5$, were observed at $\nu = 1/7$ and $1/9$
\cite{MagnetoOp}.
The FQHE states at these two filling factors have so
far not been definitively established in the more
traditional DC transport measurements that can
probe the system on a
longer length scale and a lower temperature/energy scale.
A possible explanation of this discrepancy
was proposed based on
the temperature-driven phase transitions between the
WC and the FQHE
\cite{Platzman93}. There have so far been
no reports of any FQHE-like feature at $\nu = 1/11$ with the
magneto-optical technique. Secondly, a second
luminescence line has been observed which only appears
below certain temperatures and filling factors
\cite{MagnetoOp}. It has been associated
with the formation of a solid phase. More recent time-resolved
luminescence studies seem to confirm this interpretation. Reports
have been made that even the local hexagonal order in the insulating
phase can be established \cite{MagnetoOp}. This finding is not
too surprising, and while interesting, does not address the issue of
long-range order in the system.
Recently, it has become possible to directly observe the
collective excitations of the two-dimensional electron gas
in a strong magnetic field using the inelastic light scattering
method \cite{Pinczuk}. Possible extension of this technique to studies
of the magneto-phonon dispersion in the WC would be very interesting.

All of the earlier experiments
used $n$-doped samples. However, some recent experimental work used
$p$-type doping in the
$GaAs/AlGaAS$ samples \cite{hole}.
The change in the carrier effective mass
brings a new dimension to the problem \cite{Zhu,Price93}.
Electron solidification ought to be
favored by a heavier mass at comparable
doping densities. It is indeed observed that the insulating phases
set in in these $p$-type samples around $\nu = 1/3$
\cite{hole}, compared
to $\nu = 1/5$ in $n$-doped samples. The
reentrance behavior of the insulating phases is
otherwise very similar to that
around $\nu = 1/5$
in $n$-doped samples.
A stringent consistency
check, although not a rigorous proof, of the claims that these
insulating phases are Wigner crystals, is to calculate
theoretically the solidification filling factor for the
$n$- and $p$-doped samples, and to compare with what is observed
experimentally. For this purpose, Landau-level-mixing
must be taken into account since the experimental
parameters are in a regime where $e^2 / \varepsilon l_B$ is a
factor of 10 larger than $\hbar\omega_c$, and since it is precisely
the different amount of Landau-level-mixing that gives
rise to the different transition filling factors.

Overall, there has so far accumulated a large body of suggestive
evidence of the existence of this quantum solid.
Most of the existing experimental data are consistent with,
but none have definitively established,
the formation of a Wigner crystal in these systems.
It is clear that a better theoretical
understanding of the properties of
the electron solid
under these various experimental
probes will be helpful in interpreting
in a more unified and consistent
way the existing experimental results.

We also note that an alternative explanation for the insulating phases at
$GaAs/AlGaAs$ heterojunctions discussed
here was given by Kivelson, Lee, and Zhang,
emphasizing the role of disorder \cite{CSGL}.
A generic phase diagram was constructed in terms of the
Landau-level filling factor and the strength of disorder \cite{CSGL}.
While appealing and possibly relevant to experiments
done on samples with much more disorder,
there are difficulties with this
theory when applied to experiments carried out on the best quality samples.
As we already mentioned, the insulating phases become stronger
as disorder potential is weakened in these samples, which
is difficult to understand within the framework of this
theory. There are also
some recent experimental results which are in apparent contradiction
with the predicted phase diagram \cite{Jiang1,hole}.

\subsection{Summary of our main results}

Our work can be separated into two parts: $B = 0$
and $B \ne 0$.
We summarize them in what follows.

In the case of no external magnetic field, we have studied the
variational energies of electrons localized on various
lattices, including the square lattice, the honeycomb lattice,
and the hexagonal lattice \cite{ZhuandLouie}. In investigating the lattice
dependence of the ground-state spin-ordering, we have also
studied the rectangular lattice.
The hexagonal lattice is found to have the lowest energy, in
agreement with classical Ewald energy considerations.
We also find that the square lattice
favors slightly the ferromagnetic (FM) state
whereas the honeycomb lattice favors the
antiferromagnetic (AFM) state. This is despite the fact that
both are bi-partite lattices.
The energies of different spin states on a
hexagonal lattice are very close, and beyond the
resolution of the present variational Monte Carlo calculations.
The physics of the electron solid appears
very rich even in the absence of an
external magnetic field.
We propose that the many-electron ring-exchange mechanism is
responsible for the calculated behavior \cite{Thouless,Roger,Herring}.
We will also discuss the ring-exchange processes in the
presence of a strong magnetic field, which have been
invoked as an alternative picture for the quantum Hall liquid
\cite{RingEx,Baskaran,RingExThouless,KivelsonCom}.

In the IQHE and the FQHE regime, we find that exchange
effects are unimportant in the regime of density and
magnetic field near the Wigner crystallization
point in these systems. In $n$-doped samples,
intra-Landau-level correlations are the most important,
and are well represented by the magneto-phonon wavefunction
proposed by Lam and Girvin \cite{LamGirvin}. For $p$-doped samples, however,
Landau-level-mixing is the single most important
mechanism for lowering the energy of the WC, in comparison
with exchange and intra-Landau-level correlation. However
even when $e^2/\varepsilon l_B$
is 10 times $\hbar\omega_c$, intra-Landau-level
correlations still make a very large contribution to the
total energy. Combining our WC energies with those of the
FQHE liquid with Landau-level-mixing, we find that
the solidification in $p$-doped samples will
occur around $\nu = 1/3$ \cite{Zhu,Price93}.
This is in nice agreement with the
recent experiments performed on $p$-doped samples \cite{hole}.
Effects of finite temperatures and of disorder, and
Wigner crystallization at integer filling factors will also
be discussed.

\subsection{Outline of the paper }

The balance of the paper is as follows.
In Sec.~II, we focus on the spin-ordering in the
ground-state of Wigner crystal in various 2D lattices with
no magnetic field. We interpret the results
with a picture of multi-electron
exchange interaction.
We then comment on the ring-exchange
processes in the presence of a strong magnetic field,
and examine
some overall characters of our trial wavefunctions.
We begin our investigation in
the FQHE regime in Sec.~III where we give the details of
the present VMC calculation in a strong magnetic
field. Our choice for the form of the variational wavefunction
is motivated. Results, comparison to previous work, and possible
implications for the current experiments are given in Sec.~IV.
We conclude in Sec. V.

Those who are only interested in cases with a strong magnetic field
may go to Secs.~IIC, III and IV directly.

\section{2D Wigner crystals in the absence of a magnetic field: spin-ordering }

In this section, we focus on the spin-ordering of electrons
interacting with the long-range $1/r$ Coulomb potential
on various two-dimensional lattices
with the variational
quantum Monte Carlo method.
Some comments are made for the ring-exchange
processes involving a strong magnetic
field. A general analysis of our trial wavefunction and
its possible generalization to other cases are also given.

\subsection{Results from VMC: possible role of many-electron exchange}

The spin-ordering of
electrons on several two-dimensional lattices is found to
depend strongly on the underlying lattice.
We have considered
the square
(SQ) lattice, the
honeycomb (HC) lattice and the hexagonal (HX)
lattice, in the absence of
any external magnetic field.
We focus on the first two
crystal structures because they appear to
behave very differently despite the
fact that both are bipartite.
Within our variational
wavefunctions,
the FM state is favored on the square lattice by
a much smaller margin than one by which the AFM state is favored on the
honeycomb lattice.
In addition, the rectangular lattice is studied
as a model that can continuously vary
from a square lattice to a collection of
interacting chains in 2D.
We find a transition from an FM ground-state to
an AFM ground-state as the aspect ratio deviates from one.
These results demonstrate the importance of many-particle
exchange effects for fermions interacting
with a $1/r$ potential in 2D. They also show that the
relative importance of the exchange
processes involving different number of
electrons depends strongly on the lattice geometry.

The notion of many-particle exchanges was previously invoked
\cite{Thouless,Roger}
in attempts to understand the magnetic properties of
three-dimensional
solid $^3$He. It is now well-known \cite{Thouless,Roger,PIMC,Wilkins}
that these many-particle exchanges
dominate the magnetic properties
of solid $^3$He in both 2D and 3D.
That these ring-exchanges may also affect the
magnetic properties of the
electron Wigner crystal
was
suggested by Herring \cite{Herring} in the 1960's.
Variational and Green's function Monte
Carlo calculations were
carried out previously, but only for the 2D
hexagonal lattice \cite{Ceperley}.
Two spin-orderings were considered. One is the FM state and the
other one has electrons of opposite spins aligned on
alternating chains
in the hexagonal lattice. The
latter is not a true AFM state since the
hexagonal lattice is not bipartite.
They were found to have the same
energy within the statistical noise in the calculation.
The ground-state of the spin-${1 \over 2}$ nearest-neighbor
Heisenberg antiferromagnet on a hexagonal lattice has been a subject
of tremendous amount of work \cite{HeisenHX}. To clarify the
spin-ordering of a 2D hexagonal WC is a much more demanding task still.
A semi-classical $WKB$ estimate of the various ring-exchange
frequencies
suggests that the three-particle exchange may be
more important than both
the two- and four-particle exchanges
\cite{Roger}.

The trial wavefunction used in the present VMC calculations
is of the Jastrow-Slater-type:
\begin{equation}
\psi = exp\biggl[-{1\over 2}\sum_{i\ne j}
u(\vec r_i-\vec r_j)\biggr]
D_{\uparrow} D_{\downarrow}.
\label{wfnB=0}
\end{equation}
Here $D_{\uparrow}$ and $D_{\downarrow}$ are respectively the Slater
determinants for spin-up and spin-down electrons on the lattice.
The single-particle orbitals $\phi (\vec r)$
in the Slater determinants are taken
to be in most cases
isotropic Gaussians localized about the lattice sites $\vec R_i$,
$\phi_i = e^{-(\vec r -\vec R_i)^2/r^2_G}$, except for
the rectangular lattice (see below).
The width
$r_G$ is a variational parameter to be optimized.
Additional variational
degrees of freedom in the one-particle orbitals are
introduced in further investigating the AFM state
on the square lattice.
The two-particle correlation factor in Eq.~\ref{wfnB=0}
is taken to be of the form:
\begin{equation}
u(r) = {A \over {\sqrt{r}}}
\biggl(1-e^{-\sqrt{r/F}-{1\over 2} r/F}
                            \biggr).
\label{Jastrow}
\end{equation}
It has a long range tail of $A \over {\sqrt{r}}$.
This, in the absence of the Gaussian single-particle
orbitals in the Slater determinants, yields
a longitudinal phonon
dispersion $\propto q^{1/2}$ for small-$q$ which is
a result of the Coulomb interaction in 2D.
In the limit of $r \rightarrow 0$,
${du \over dr} = -{A \over 3} ({1\over F})^{3/2}$ and
$u=A/\sqrt{F}$.
With properly chosen A and F, the short
range cusp-condition \cite{cuspcond}
can be satisfied.
We have determined these two parameters in our trial wavefunction
variationally. The
optimal values are always very close to those
given by the cusp-conditions and the long wavelength
longitudinal phonon considerations. Calculations
are done using the Metropolis algorithm
with periodic boundary conditions in 2D.

Our results for the hexagonal structure are the same as
those of existing VMC calculations reported previously where
a somewhat
different Jastrow factor was used \cite{Ceperley}.
In Table~I,
we show a comparison between the present VMC calculation and
a previous VMC calculation \cite{Ceperley}
for a simulation cell containing 56
electrons. Finite size effects have been examined
using different size simulation cells. To the significant
digits given in Table~I, they are negligible.
For reference, results from a fixed-node Green's
function Monte Carlo calculation \cite{Ceperley}
are also shown. As mentioned above,
because of the frustration effects,
the energy differences between different spin-orderings of the WC
on the hexagonal lattice are too
small to be studied with the present method.
{}From now on, in discussing the spin structures, we will restrict
ourselves to the SQ lattice and the HC lattice.

The energies here are dominated by the classical Ewald energy.
For the three lattices studied here,
they are respectively (in atomic units):
$E_{Ewald}=-1.10610/r_s$ for the HX lattice,
$E_{Ewald}=-1.10024/r_s$ for the
SQ lattice, and $E_{Ewald}=-1.06841/r_s$
for the HC lattice.
At a given density, the HC lattice has the smallest
nearest-neighbor distance since it is the least close-packed among
the three, hence it has the highest classical energy.
We find that quantum effects tend to reduce the energy differences,
but they are not large
enough to reverse the ordering.
In Table~II, we show, for these three lattices, the VMC energies
calculated at $r_s = 30$, 50, 70, and 100, all with
complete spin polarization. Again, finite size effects
are negligible compared to the significant digits given.

We now focus on the different spin-orderings
in the SQ and HC lattices.
In the present work, the FM state is formed with all
the single-particle orbitals having the same spin. Thus the total
wavefunction contains only one Slater determinant.
The AFM state
on these bipartite lattices is
constructed by occupying the single-particle orbitals on
one sublattice with spin-up electrons
and the other with spin-down electrons. The spin-dependent
cusp-conditions \cite{cuspcond} are satisfied with a spin-dependent
Jastrow factor.
In Figs.~1 and 2, we show the energy differences between the FM state
and the AFM state on the SQ lattice and the HC lattice for several
$r_s$.
As we can see, the FM state is slightly lower
in energy than the AFM on the square lattice. But on the honeycomb
lattice the AFM is lower in energy by a
substantial margin for $r_s \leq 100$.
Table~III illustrates that the finite size effects at
$N \sim 50$ are already negligible.
Results reported in the figures
are calculated with the largest simulation cells shown in Table~III.

In general, it is more difficult to construct
a good trial wavefunction for a more disordered state.
In the present case, one might suspect that the
FM state is slightly favored over the AFM state in the
variational calculations. Of course, this will not change
our conclusion regarding the HC lattice, where the more
disordered phase (AFM state)
already has a lower energy. For the SQ lattice, we have
made the following two attempts to lower the energy of
the AFM state:

\noindent
1). Instead of having Gaussians localized on a single lattice site
in the one-particle orbitals, we have used:
\begin{equation}
\phi_j(\vec r) = e^{-(\vec r - \vec R_j)^2 / r_G^2 } + C
\sum_{i=nn} e^{-(\vec r - \vec R_i)^2 / r_G^2},
\end{equation}
with proper normalization factors.
Here $C$ is a variational parameter and $\sum_{i=nn}$ indicates
summing over the nearest neighbor sites. This
form of $\phi(\vec r)$
distributes some weight of the one-particle orbital onto
its neighboring sites. Noting that the nearest
neighbors have the opposite spin, we expect that the sublattice
magnetization of the 2D antiferromagnet
could be adjusted by changing $C$.

\noindent
2). We have also tried one-particle
orbitals similar to the above, but
with some weight on the hollow sites in the lattice rather than
on the neighboring lattice sites.
The motivation was to lower the kinetic energy at
minimal cost to the
interaction energy.

In both of these cases, the AFM state energy on the SQ lattice was not
lowered to within our statistical accuracy.
Despite these efforts, we
can not be absolutely
certain that the FM state is lower in energy
than the AFM state on the SQ lattice.
In fact, we do not know if the
ground-state of a WC on a SQ lattice would be magnetically
ordered at all.
The closeness of these two states on a SQ lattice
leaves open the possibility that the ground-state on
a square lattice may have a more subtle magnetic ordering
than what we are able to examine in the present variational
calculations.
It is safe to conclude, however,
that our calculations illustrate clearly
the
qualitative difference between the bipartite square and honeycomb
lattices in their ground-state spin-ordering.

\subsection{Discussions and further tests}

For fermions on a lattice, it has been
argued by Herring \cite{Herring}
and by Thouless \cite{Thouless} that, in general,
ferromagnetism is favored by ring-exchange processes involving
odd number of particles ($(2n+1)$-exchange),
whereas antiferromagnetism is favored by
ring-exchange processes involving even number of particles
($2n$-exchange).
A heuristic argument may go as follows.
Take the example of an arbitrary $(2n+1)$-ring-exchange.
A ring-exchange of $(2n+1)$-particles is
an even permutation. Thus the total
wavefunction must keep the same sign upon this ring-exchange.
To minimize the energy,
one would like to keep the number of nodes in the spatial
part of the wavefunction minimal.
Thus, a wavefunction
in which the spatial part is
totally symmetric will be favored. We then must make the
spin part of the wavefunction totally
symmetric as well.
Consequently, the FM state, where all spins are aligned,
will be favored by such exchanges.
Similarly one concludes that
AFM is favored by $2n$-exchanges. Following this line of reasoning,
our results suggest that
two-particle exchanges are more important on the
HC lattice while three-particle exchanges are relatively
more important on the SQ lattice. The competition between
them determines the ground-state spin-ordering.

Compared with the interaction potential
in solid $^3$He \cite{Roger}, the
$1/r$ Coulomb interaction
is much softer for small $r$.
This results in lesser steric constraint in exchanges
involving fewer particles.
One might therefore speculate that exchange frequencies
as a function of $n$ will decay faster
for electrons on 2D lattices than for solid $^3$He
\cite{Roger}.
These plausibility arguments could be checked by either
the semi-classical $WKB$ approach \cite{Roger}
or by using the
path-integral Monte Carlo approach \cite{PIMC} which has been
applied to study the ring-exchange
frequencies in body-centered-cubic
solid $^3$He.

Although the variational Monte Carlo method was the first
\cite{Wilkins} to be used
in studying numerically the
four-particle exchange frequency in
solid $^3$He, it has later been shown to be
quantitatively unreliable \cite{Roger,PIMC}.
This is largely due to the fact that in the case of
solid $^3$He the exchange energies are only $10^{-4}th$
of the typical phonon energies. Thus an excellent
variational wavefunction
for the total energy might be inadequate
for estimating the
exchange frequencies.
In the present case however,
the energy differences between the AFM and the FM
spin-orderings at $r_s = 30$ are about $1\%$ for SQ and $5\%$ for HC
of the total zero-point-motion energy.
In fact
for HC lattice, $E_{FM} - E_{AFM}$ at this $r_s$ approaches
the total energy difference between the SQ
lattice and the HC lattice. We therefore believe
that the qualitative difference in the ground-state
spin-ordering between these two bipartite lattices
is not an artifact of the variational method.

The energy splitting between the
FM state and the AFM state is
found to be very sensitive to the
Gaussian width $r_G$ in the one-particle
orbital. In Fig.~3, we show this sensitivity for the HC lattice.
While the sharp drop on the small $r_G$ side may be
attributed to the exponential decrease of the exchange overlap
integral, the somewhat slower drop on the large $r_G$ side
could be due to the different $r_G$-dependences
of the two-particle and three-particle
ring-exchange processes. The energy
splitting peaks at an $r_G$
smaller than one which optimizes the energies.

Further support for the importance of the multi-particle
ring-exchanges for electrons on a 2D lattice
is given by studying the rectangular
lattice as it varies from a square lattice
to a set of weakly interacting chains.
Here, we have used anisotropic Gaussians in accordance with the
aspect ratios. The sizes of the Gaussians
along different directions are kept proportional to the
two side lengths of the rectangle.
In the essentially one-dimensional limit of a set of
weakly interacting chains, two-particle exchange
is expected to dominate. This is because multi-particle
exchanges will involve tunneling over longer distances with the
same or higher potential barriers than the two-particle
exchange in this limiting
case. Consequently the AFM state should have a lower energy if the aspect
ratio is large enough. In Fig.~4, we show the energy difference
between the FM state and the AFM state on a rectangular lattice
as a function of its aspect ratio
for $r_s=30$. Indeed, we see a transition in
the region of $a/b = 1.15$ to $a/b = 1.20$.
All of these observations from our calculations are consistent with
the multi-electron ring-exchange picture. We thus believe that the
ring-exchange processes play an important role in
determining the ground-state spin structure for
electrons on
these 2D lattices.

\subsection{Many-particle exchanges in the Wigner crystal in a strong
magnetic field }

It is of interest to ask how the
presence of a strong magnetic field may affect
the many-particle exchange processes.
In two-dimensions,
at the mean-field-theory level, at suitable magnetic
filling factors, the physical magnetic field may be gauged
away \cite{CSGL}.
One is then left with composite particles which carry
flux quanta moving in an effective
zero magnetic
field. Since statistics of the particles may be altered in such
formal transformations of the Hamiltonian, and since
the exchange interaction is intimately related to the
statistics, it is expected that
the many-particle exchange effects may be dramatically affected by
a magnetic field.

This issue has been taken up in a series of papers by
several authors in which they argued at filling factors
such as $\nu = 1/3$, $2n$ and $(2n+1)$-exchanges will add
coherently and yield a condensation of the large ring-exchange
processes \cite{RingEx,Baskaran}.
This notion as
it was first proposed \cite{RingEx},
while appealing physically, cannot survive
the large magnetic field limit on a lattice.
In the resultant localized crystal phase,
the exchange interactions can be
made arbitrarily small.
Our numerical calculations
also show clearly that exchange processes are unimportant
compared to the correlation effects in a
magnetically induced Wigner crystal.
Within our numerical resolution, the total energy
changes smoothly with the applied magnetic field,
or with the magnetic filling factor $\nu$, showing
no downward cusps necessary for the experimentally
observed incompressibility at odd-denominator  filling factors.

This problem of invoking the lattice
is circumvented later by Baskaran and Lee, Baskaran,
and Kivelson \cite{Baskaran}
who showed that the crystal phase is
not necessary for such a condensation to occur, although it is
convenient for the sake of visualization. There is also a
formal mapping between the density matrix from
the Laughlin wavefunction and that from the
condensed ring-exchange processes \cite{Baskaran}.

Thouless and Li have raised another criticism, regarding the sign
of the ring-exchanges at filling factors that are
the inverse of odd integers \cite{RingExThouless}.
We repeat their argument for
spin-polarized systems with
the following diagram (Fig.~5). The uparrows
(the downarrows) indicate  that the energy goes up (down)
when such exchanges are included.
One might then conclude that
the ring-exchange processes at odd-denominator filling
factors would in fact make them energetically less
favored compared to its immediate neighboring filling factors.
Therefore, the energy will show an upward cusp at odd-denominator
filling factors, instead of the downward cusp as implied by the
observation of the FQHE \cite{FQHE0,Laughlin}.

We think that this argument is
in fact not valid in the FQHE regime \cite{KivelsonCom} because
it rests on the assumption
that the energy always goes up as the number
of nodes in the spatial part of the wavefunction increases.
We argue below that existing calculations explicitly
show that this is not the case
for the very system of interest here.

We make use of a recent work by
Xie, He, and Das Sarma where they considered
two systems of identical interacting particles on
a sphere \cite{XieHeDS}.
The only difference between the two is that
one is bosonic and the other
fermionic.
In the absence of a magnetic field,
the Bose system will in general have a lower energy
\cite{Feynman}.
But in the presence of magnetic field,
the urge to lower the interaction energy and the possibility of
doing so without costing kinetic energy make it under
certain circumstances favorable
to have nodes in the many-body wavefunction.
This is indeed what was found
from the exact diagonalization of the few
particle system \cite{XieHeDS}: At filling factors $\nu = 1/2$,
the bosonic system has a lower energy but at $\nu = 1/3$ the
fermionic system has a lower energy for Coulomb interactions.
The qualitative trend appears to continue for all the
filling factors examined \cite{XieHeDS}.
As the magnetic field decreases or as the filling factor increases to
well above 1, eventually the Bose system will have a lower energy.

\subsection{Qualitative features of the trial wavefunction }

In order to gain a more physical understanding about
what kind of correlations are included in our trial wavefunction,
it is instructive to transform our wavefunction in terms of
the phonon coordinates. This transformation cannot be carried out
in the most general case, similar to earlier work on the liquid
phase \cite{Edwards}. We therefore ignore exchange, {\it i.e.},
approximate the Slater determinants $D_{\uparrow}D_{\downarrow}$
with:
\begin{equation}
D_{\uparrow}D_{\downarrow}
= e^{-\sum_{i = 1}^N(\vec r_i - \vec R_i)^2/r_G^2}.
\label{DupDdown}
\end{equation}
Using:
\begin{equation}
{1\over 2}\sum_{i\ne j}u(r_{ij})
= {1\over 2N}\sum_{\vec q}u(q) [\rho_{\vec q}\rho_{-\vec q} - N],
\end{equation}
where
\begin{equation}
\rho_{\vec q} = \sum_{j = 1}^N e^{i\vec q \cdot \vec r_j},
\end{equation}
we can rewrite our original wavefunction, Eq.~\ref{wfnB=0}, as:
\begin{eqnarray}
\psi_T(\vec r_1,\vec r_2,...,\vec r_N)
&=& e^{
   -{1\over 2}\sum_{\vec q} u(q)[\rho_{\vec q}\rho_{-\vec q} -N]/N
   -\sum_{i=1}^N(\vec r_i - \vec R_i)^2/r^2_G
    } \nonumber \\
&=& e^{K},
\end{eqnarray}
which defines $K$.

Under the assumption that the deviations of $\vec r_i$ from
$\vec R_i$ are small, we may expand $\rho_{\vec q}$ to first
order in $\vec \xi_i = \vec r_i - \vec R_i$, and obtain for $K$:
\begin{equation}
K =  -{1\over 2} \sum_{\vec k} \vec \xi_{\vec k} \cdot (\vec k u(k) \vec k
+ 2 {\bf I}/r^2_G) \cdot \vec \xi_{-\vec k}
\end{equation}
where ${\bf I}$ is the unit matrix and
the phonon coordinates $\vec \xi_{\vec q}$ are
defined as:
\begin{equation}
\vec \xi_{\vec q} = {1\over \sqrt{N}} \sum_{i=1}^N \vec \xi_i
e^{-i\vec q\cdot \vec R_i}.
\end{equation}

This form is simply a product of $N$-noninteracting
simple harmonic oscillators. The kernal $\vec k u(k) \vec k +
2{\bf I} / r^2_G$ can be diagonalized, yielding:
\begin{equation}
\omega_L = 2/r_G^2 + u(k) k^2,
\end{equation}
and
\begin{equation}
\omega_T = 2/r_G^2,
\end{equation}
where $\omega_L$ is the eigenfrequency of the longitudinal
mode from the diagonalization of the tensor, and
$\omega_T$ is the transverse mode frequency. In $D$-dimensions,
there are $D-1$ degenerate transverse modes from this simple analysis.

The physics of this line of reasoning is quite clear. The
isotropic Gaussians are by themselves $N$ uncoupled
$D$-dimensional isotropic harmonic ocsillators, giving
a finite frequency to both the transverse and the longitudinal
modes. The correlation factor $u(r)$ in its present form
is only a function of inter-particle distance, thus only
affects the density fluctuations in the long-wavelength
limit where our phonon expansion is valid. This is precisely
why $u(r)$ only enters the frequency of the longitudinal mode
which couples to density fluctuations.

It is known, for Coulomb interacting systems,
long wavelength density fluctations have a finite frequency
in 3D while in 2D it disperses as $\sqrt{q}$. These statements
regarding the longitudinal fluctuations are
true for both the solid and the liquid phases. They
may be used to determine the asymptotic behavior of the
optimal $u(r)$ used in the variational calculations.
We note that the $\sqrt{q}$-dispersion for the
longitudinal phonon in 2D is violated in the long-wavelength
limit by our variational wavefunction. Similarly, the
linear-in-$q$ dispersion of the transverse phonon is also
violated. This observation suggests that, while our
wavefunctions may be very accurate for ground-state energies,
it can not be used directly for calculating phonon frequencies
in the small $q$-limit. We remark that similar violations
of the long-wavelength phonon dispersions in electron Wigner crystals
with no external magnetic fields,
and in helium solids also appear in nearly all of the previous
variational Monte Carlo calculations.

Not only does this transformation provide some insight into
our trail wavefunction, it can also be used constructively
to find an
optimal $u(k)$ in certain cases. In the liquid phase, the
approximation equivalent to the one we made above for the solid
is the random-phase-approximation for the total energy,
which is then minimized with respect to
$u(\vec k)$
\cite{Edwards}.
As an illustration
of the general principle, we have also applied this idea to the
problem of bcc solid hydrogen in the
Mott insulating regime and derived an optimal electron-electron
correlation factor. In this case, we find:
\begin{equation}
2u(k) = - 1 + \sqrt{ 1 + {4m \over \hbar^2k^2}
(v_k + 2 \vec k \cdot {\bf s} \cdot \vec k/k^4)},
\end{equation}
where
\begin{equation}
{\bf s} =  - {e^2 \over 2} \sum_{j \ne 0}
{ {3(\vec R_j - \vec R_0)(\vec R_j - \vec R_0)
- {\bf I}(\vec R_j - \vec R_0)^2} \over
{|\vec R_j - \vec R_0|^5}
},
\end{equation}
and the Fourier transform of the Coulomb interaction in 3D is:
\begin{equation}
v_k = 4\pi e^2 / k^2.
\end{equation}
Here $\vec R_0$ is the position of an arbitrary
proton in the bcc lattice and $\vec R_j$ are the
positions of all the protons.
In the absence of the proton lattice structure,
which amounts to simply setting ${\bf s}$=0 as one can verify
in the course of derivation,
we have
\begin{equation}
2u(k) = -1 + \sqrt{1 + 4mv_k/\hbar^2 k^2}.
\end{equation}
The problem is reduced to that of a jellium and
the above result agrees with
earlier work for this quantity \cite{Edwards}.

\section{Energy of a Wigner crystal in FQHE regime: Method }

In the rest of this paper, we focus on cases where a large magnetic
field is involved.
Effects of exchange, intra-Landau-level correlation,
and Landau-level-mixing on the total energy, and their dependence on
the carrier mass and magnetic field strength are
examined on equal footing.
In Sec.~IIIA, we briefly review the problem.
In Secs.~IIIB and IIIC, we describe the Hamiltonian
and the variational wavefunctions used, and the technical details
in their evaluation along the Monte Carlo walks.
In Sec.~IIID, we assess the quality
of the trial wavefunction.
In this section, we focus more on the qualitative aspects of our
calculations.
Numerical results and discussions are presented in Sec.~IV.

\subsection{Brief overview }

There have been many studies on $B$-field induced Wigner crystals
in 2D. Most of these were carried out within the Hartree-Fock
approximation \cite{WCHF,LLMHF}.
An alternative approach has been to expand the
Hamiltonian in terms of phonon coordinates and to seek to perturbatively
improve the results \cite{Chui91,ChuiOddTerms}.
The former is variational but contains no information
regarding the crucial correlation effects. The latter is no longer
variational and existing calculations show that the rate of convergence is
unsatisfactory.
With a few exceptions \cite{Zhu,LLMHF,Yoshioka84},
only the lowest Landau-level states were
considered.

The present approach has the advantage of being variational.
In addition, by varying the trial wavefunctions used,
we are able to obtain a quantitative
understanding of the roles played by exchange, intra-Landau-level
correlation, and Landau-level-mixing in a Wigner crystal. Special
attention is paid to the interplay between these effects
and the experimental parameters: the carrier density, carrier mass,
and the strength of the magnetic field.
{}From our calculations, we find that the
effects of Landau-level-mixing are indeed large enough to account for the
observed
difference in $\nu_c$ between the eletron and the hole
$GaAs/AlGaAs$ systems.
For $GaAs$, the relevant materials parameters are:
$\varepsilon = 13$, electron effective
mass $m_e^* = 0.067 m_e$, and heavy-hole effective mass
$m_h^* = 0.35 m_e$. Experimentally for the present heterostructure,
the heavy-hole effective mass is less certain \cite{Stormer83}.
For the present heterojunction system
in the strong magnetic field limit, we only need to consider the
heavy-hole band for $p$-doped samples.
We now describe details of the
present VMC calculations involving
a strong magnetic field.

\subsection{Present VMC calculations: the Hamiltonian}

The exact Hamiltonian for 2D electrons of
effective mass $m^*$ in a magnetic
field is given by:
\begin{equation}
H = \sum_i {{(\vec p_i + e\vec A(\vec r_i))^2} \over {2m^*}}
+ {e^2 \over {2\varepsilon}} \sum_{j\ne i} {1 \over {r_{ij}}}.
\label{HamilB}
\end{equation}
Here $\vec A$ is the vector potential,
$\vec A = (-yB/2, xB/2)$ in symmetric gauge
with $\vec B$ in $+z$ direction.
$\varepsilon$ is the dielectric constant
of the host material (${e^2 \over \varepsilon} = m^* = 1$
in effective atomic units).
The Zeeman term is left out of
Eq.~\ref{HamilB} since we assume total spin-polarization.

In choosing a particular
gauge for the vector potential $\vec A$
in the Hamiltonian, we must also choose an origin, which
breaks the continuous translational invariance.
As a result of the generalized periodic boundary conditions \cite{Zak},
all the rational fields can be studied directly
in our numerical work. Properties of the WC at irrational fields can only
be obtained to the extent continuity holds for the particular
physical quantity.
In Sec.~IVF, we provide numerical evidence that
there is no cusp
in the WC total energy as a function of filling
factor.

A finite
simulation cell with modified periodic boundary conditions \cite{Zak}
is used, and only
the hexagonal lattice is
considered in view of the results presented in Sec.~II.
With the kinetic energy quenched by
the magnetic field, it is expected that the
hexagonal structure would be made even more stable (compared
to the $B = 0$ case) than
the other 2D lattices.

The evaluation of the $1/r$ interaction energy at each step of
the Monte Carlo walk is
not affected by the magnetic field. It is done with the usual
Ewald sum method in 2D at each step of the Monte Carlo walk.
To evaluate the kinetic energy, let us first define:
\begin{mathletters}
\begin{equation}
u = x + iy,
\end{equation}
\begin{equation}
v = x - iy.
\end{equation}
\end{mathletters}
The kinetic energy operator is now:
\begin{eqnarray}
K.E.
&=& \sum_i k.e.^{(i)} \nonumber\\
&=& {1 \over {2m^*}}\sum_i {(\vec p_i + e\vec A(\vec r_i))^2} \nonumber\\
&=& {1 \over {2m^*}}\sum_i
                  \biggl\{ {-4{{\partial^2} \over {\partial u_i\partial v_i}}
             +{1 \over {l_B^2}}(u_i{\partial \over {\partial u_i}}
                               -v_i{\partial \over {\partial v_i}})
             +{u_iv_i \over {4l_B^4}}
                           }
                  \biggr\}.
\label{ke0}
\end{eqnarray}
The local kinetic energy
\begin{equation}
K.E._{loc} = {1 \over \psi}
\sum_i {1 \over {2m^*}} {(\vec p_i + e\vec A(\vec r_i))^2} \psi,
\label{ke1}
\end{equation}
is transformed into the following form:
\begin{eqnarray}
{2m^* } K.E._{loc} & = &
-4(J_{uv} + D_{uv})
-4(J_u \cdot J_v + J_u \cdot D_v + D_u \cdot J_v) \nonumber \\
                          &   & \mbox{}
+{1\over l_B^2}(u\cdot D_u - v \cdot D_v)
+{1\over l_B^2}(u\cdot J_u - v \cdot J_v)
+{uv \over 4l_B^4},
\label{ke}
\end{eqnarray}
where the many-body wavefunction is written as
\begin{equation}
\psi (\vec r_1,\vec r_2,...,\vec r_N)
=J(\vec r_1,\vec r_2,...,\vec r_N) \cdot
 D(\vec r_1,\vec r_2,...,\vec r_N),
\end{equation}
and,
\begin{mathletters}
\begin{equation}
J_u = {{\partial ln J} \over {\partial u}},
\end{equation}
\begin{equation}
J_v = {{\partial ln J} \over {\partial v}},
\end{equation}
\begin{equation}
J_{uv} = {{\partial^2 ln J} \over {\partial u\partial v}},
\end{equation}
\begin{equation}
D_u = {{\partial ln D} \over {\partial u}},
\end{equation}
\begin{equation}
D_v = {{\partial ln D} \over {\partial v}},
\end{equation}
\begin{eqnarray}
D_{uv} &=& {{\partial^2 ln D} \over {\partial u\partial v}}
      + D_u \cdot D_v \nonumber \\
       &=& {1 \over D}{{\partial^2 D} \over {\partial u\partial v}}.
\end{eqnarray}
\end{mathletters}
The magnetic length $l_B$ is $\sqrt{\hbar c /eB}$ as before.
We have suppressed the electron index $i$ in all the terms on the
right-hand-side of Eq.~\ref{ke}.
Each of them is collected along the walk and used to
construct the local kinetic energy at a given configuration. The above
transformation is not necessary --- one can in fact just as easily
evaluate the local kinetic energy in terms derivatives with respect
to $x$ and $y$, although it does make the distinction between
lowest Landau-level states and the higher Landau-level states more
apparent. This is because the lowest Landau-level wavefunctions
can be written as a product of a Gaussian with
an analytic function of $u$. In other words, a many-body wavefunction
restricted to the
lowest Landau-level states will not contain any
$v$-dependence in its Jastrow factor $J$, that is, $J_v = J_{uv} =0$.

Given a trial wavefunction, we
sample the total energy with the Metropolis scheme \cite{VMC}.
The fairly strong localization
of the electron wavefunction
makes it possible for us to totally eliminate the finite
size effects. (The fact that we are dealing with
a 2D systems also reduces the total number of particles
needed for convergent results compared with
a 3D system of the same linear size.)
Most of our
numerical calculations are carried out for a simulation cell
with 100 spin-aligned electrons.
Calculations with different size simulation cells
show that the resulting finite size effects are smaller than the
statistical noise in our results. We will give more details
regarding finite size convergence.
No finite size scaling is needed to
extrapolate the energy to the thermodynamic limit.
The numerical aspects of the calculation are therefore
well under control.

\subsection{Many-body wavefunction}

Denoting the simulation cell by vectors $\vec L_x \times \vec L_y$,
we may write
the single-particle orbital that is localized
about the lattice site $\vec R_j$ and satisfies the
generalized periodic boundary conditions as:
\begin{equation}
\phi_j(\vec r) = {1 \over \sqrt{2\pi}} {\beta \over l_B}
\sum_{\vec T} exp \biggl\{ - {\beta^2 \over 4l_B^2}
   (\vec r - \vec R_j - \vec T)^2
   + {i \over 2l_B^2}\bigl[\vec r \times \vec R_j +
                           \vec r \times \vec T   +
                           \vec R_j \times \vec T        \bigr]_z
                  \biggr\}.
\label{wfn}
\end{equation}
Here $\vec T = n_x \vec L_x + n_y \vec L_y$ with arbitrary
integers $n_x$ and $n_y$.
$\beta$ is
a variational parameter which determines the
localization of the wavefunction. Changing
$\beta$ does not affect the phase factors appearing in
$\phi(\vec r)$.
(For $\beta = 1$, $\phi(\vec r)$
lies entirely within the lowest Landau-level.)

One may form either a Slater determinant or a simple product
of these single-particle orbitals.
When multiplied by a
purely periodic Jastrow factor, both of the
resultant many-body wavefunctions satisfy the generalized
periodic boundary condition in a vector potential \cite{Zak}.
Without the Jastrow
factor, none contains correlation but the former does contain
exchange. Therefore, we can
obtain a rigorous upper bound for the size of
exchange energy using these wavefunctions.

We now motivate the Jastrow factor that we use.
It consists of two parts: One is the same as that
in the absence of the magnetic field which is shown to
be quite accurate in that case; the second part,
arising from the zero-point-motion of the magneto-phonons, is
peculiar to cases involving a strong magnetic field.
The latter is found to be more important in terms of its effects
on energy. Of course, such a separation is not entirely strict.

Our derivation for the magneto-phonon correlated
wavefunction is slightly different
from Lam and Girvin's work \cite{LamGirvin}.
They adopted the lowest
Landau-level approximation at the outset and sought to optimize the
harmonic Hamiltonian within the lowest Landau-level
sub-Hilbert space. We have chozen to solve the harmonic Hamiltonian
exactly without making the lowest Landau-level approximation.
We can then obtain the Lam-Girvin lowest Landau-level magneto-phonon
wavefunction by taking the large magnetic field limit. Clearly these
two procedures are equivalent in
the strong-field limit, although our scheme is somewhat more
flexible as it may be used to include some of the Landau-level-mixing
effects. Our derivation closely parallels that of Chui and
his coworkers' \cite{Chui91}, although they have focused more
on the eigenvalues than on the
wavefunctions.

After we take the large field limit, our wavefunction
is the same as that given by Lam and Girvin \cite{LamGirvin}.
Using their notation, the
wavefunction $\psi_{cor}$ restricted to the lowest
Landau-level is:
\begin{equation}
\psi_{cor} =
exp { \biggl[ {A_p \over 4l_B^2} \sum_{i,j} \xi_i B_{ij} \xi_j \biggr] }
{}~\prod_i \phi_i(\vec r_i),
\label{LGwfn}
\end{equation}
with $A_p = 1$.
$\phi_i$'s are
the single-particle orbitals in Eq.~\ref{wfn} without
the $\vec T$'s. $\xi_i$ is the displacement
of the $ith$-electron from the lattice site $R_i$, written
in complex coordinates.
The $B_{ij}$'s, appearing in the Jastrow factor,
couple the motion of the $ith$-electron
with that of the $jth$-electron.
Its Fourier transform $B(\vec k)$ is:
\begin{equation}
B(\vec k) = e^{i\theta_{\vec k}} { {\omega^0_L(\vec k) - \omega^0_T(\vec k)}
                      \over {\omega^0_L(\vec k) + \omega^0_T(\vec k)} },
\label{Bij}
\end{equation}
with,
\begin{mathletters}
\label{thetak}
\begin{equation}
cos \theta_{\vec k} = -{{(D_{xx}-D_{yy})/2}
\over {\sqrt{(D_{xx}-D_{yy})^2/4+D_{xy} \cdot D_{yx}}}},
\end{equation}
\begin{equation}
sin \theta_{\vec k} =
-{ D_{xy} \over {\sqrt{(D_{xx}-D_{yy})^2/4+D_{xy} \cdot D_{yx}}}}.
\end{equation}
\end{mathletters}
All quantities in Eqs.~\ref{Bij} and \ref{thetak}
are those of a hexagonal WC
in zero fields: $D_{xx}$ and $D_{yy}$ are the diagonal elements,
$D_{xy} = D_{yx}$ the off-diagonal elements,
of the dynamical matrix at $\vec k$; and $\omega_T^0$
and $\omega_L^0$ are
the transverse and longitudinal phonon frequencies
\cite{Bonsall}.
We emphasize that this wavefunction is for distinguishable particles
correlated within the lowest Landau-level.

In our variational calculations, we have allowed the coefficient
of the overall
exponent in Eq.~\ref{LGwfn}, $A_p$,
to vary from 1. This is found
to have only small effects on the total energies for $A_p \sim 1\pm
0.2$. The optimal $A_p$ is found to be very close to 1 (see
discussions below).

One comment is in order here.
When we adopt the supercell geometry,
the $\vec k$'s in $B(\vec k)$ are those compatible
with the periodic boundary conditions. This is because the
Fourier transform of $\xi_i$ in different simulation cells only
has components at these selected $\vec k$'s.
Let us consider an arbitrary electron, say, electron 1
at $\vec r_1$ localized around  a lattice site $\vec R_1$. Its motion
is correlated with that of electron $\vec r_2$ localized
around $\vec R_2$, in the form of $\xi_1 B(\vec R_1,\vec R_2)\xi_2$.
Due to the periodic boudary conditions, it is also correlated with
all the images of $\vec r_2$ in a repeatative fashion. Physically,
this correlation must take the form of
$\xi_1 B(\vec R_1, \vec R_2+\vec T)\xi_2$, not
$\xi_1 B(\vec R_1, \vec R_2) (\xi_2 + complex (\vec T))$.
This procedure is
precisely the same as that in Lam and Girvin's work \cite{LamGirvin}
when they adopt the
special $\vec k$-point sampling scheme to evaluate the energy and
the wavefunction.
Therefore, the magneto-phonon correlation factor
is not periodic in $\vec r_i$, but rather
in $\xi_i$. A technical point related to this physical requirement is that
each electron must now be associated with a particular $\vec T$, in
addition to an $\vec R$, if we wish to use this magneto-phonon correlation
factor.
In the magneto-phonon picture, all electrons are
distinguishable, including those at different $\vec R$'s or
at different $\vec T$'s. For two electrons within the same $\vec T$, we
always evaluate the $\xi$'s from the fixed $\vec R$'s even though at a
given step of the Monte Carlo walk, $\vec r_1$ may be closer
to $\vec R_2$ than to $\vec R_1$. Same is true for
two electrons with the same label 1 but lie in different $\vec T$'s.
This situation is entirely analogous to the evaluation of the
total phonon zero-point-motion energy in a semiconductor,
where all ions are distinguishable, but move in unison
from one supercell to another.

The above correlation factor by construction does not
mix in higher Landau-levels.
We have also used a Landau-level-mixing correlation
factor that has the same form as that in the absence of the
magnetic field. For ease of
reading and discussion, we rewrite it here:
\begin{equation}
u(r) = {A \over {\sqrt{r}}}
\biggl(1-e^{-\sqrt{r/F}-{1\over 2} r/F}
                            \biggr),
\label{Jastrow1}
\end{equation}
where $r$
is the distance between the two electrons.
When we adopt this correlation factor, electrons are
not attached to specific lattice sites or to
a particular simulation cell.
We can therefore combine this correlation factor
with the Slater determinant formed by one-particle orbitals in
Eq.~\ref{wfn}. The resultant wavefunction
is one in which all electrons are identical regardless of the
$\vec R$'s and $\vec T$'s that they happen to lie close to
at any particular moment.
This allows us to assess the effects of exchange quantitatively.
The correlation factor in Eq.~\ref{Jastrow1} mixes even
and odd terms in the sense of
phonon expansion of the Hamiltonian \cite{Chui91},
and allows one to adjust the cusp-condition. In all
the calculations reported below, we have used
$A/(3F^{3/2}) = 1/3$. We have tested the cusp-condition with
other values and found $A/(3F^{3/2}) = 1/3$ to be near the optimal
in all cases. We give the details in the next section.
We note that the Jastrow factor in Eq.~\ref{Jastrow1}
also modifies
the intra-Landal-level correlation. This can be seen by
projecting the wavefunction with Eq.~\ref{Jastrow1} onto the lowest
Landau-level
subspace.

The evaluation of $1 \over \sqrt{r}$ in the Jastrow
factor is done with
the usual Ewald sum method, now in 2D. Its various derivatives, needed in
the kinetic energy calculation, are treated in the same way.

The magneto-phonon correlation factor is expected to be quite good
for long-range correlation effects. We
also expect the $1 \over \sqrt{r}$-term
to be reasonable for short and
intermediate-range correlations. It is near optimal in the absence
of the magnetic field, as can be seen from comparison with
GFMC calculations (see Table~I).
The combined result of these
two correlation factors interpolates smoothly between the strong-field
and the weak-field limits. It is thus expected to yield an excellent
correlated wavefunction for a Wigner crystal in a strong magnetic
field.

\subsection{Quality of the variational wavefunction: comparison
to fixed-phase quantum Monte Carlo results}

After our work \cite{Zhu} was published,
another theoretical work appeared \cite{Ortiz} in which an extension of the
fixed-node Monte Carlo method to systems without time-reversal
symmetry was applied to the 2D electron system in a strong
magnetic field. We now compare our variational Monte Carlo
results to their work. The ``fixed-phase" method is, in
principle, able to find the lowest possible energy for a given
choice of the phase of the wavefunction. It should be noted that the phase
of the wavefunction
must be fixed for every point in the entire configuration space.

Two useful conclusions emerge from the comparison between
the present variational Monte Carlo results and the fixed-phase
diffusion Monte Carlo results:

\noindent
I) By intentionally restricting the variational freedom in
our trial wavefunction so that its phase is the same as that
used in Ref.~\cite{Ortiz}, the variational method is able to reproduce the
diffusion Monte Carlo energy to within the accuracy of the published results.

\noindent
II) As a result, the source of the relatively
poor quality of the choice of the phase for the Wigner
crystal wavefunction used in Ref.~\cite{Ortiz} becomes apparent:
It is due to the fact that the magneto-phonon correlations
are not sufficiently included by the choice of the phase.
The energy of the Wigner crystals given by
the fixed-phase diffusion Monte Carlo in Ref.~\cite{Ortiz}
can be lowered by
choosing a phase including the magneto-phonon correlations \cite{footnote1}.
However, in view of
I), it is expected that the
present variational Monte Carlo results for the WC
including the magneto-phonon
correlation effects are sufficient for calculating the FQHE-WC phase
boundary.

We now give the numerical details of the comparison. The phase
of the WC wavefunction used in Ref.~\cite{Ortiz}
will be the same as that used in this work if
we set $A_p = 0$, that is, if we turn off the magneto-phonon
correlations intentionally. We do so, and then optimize
the energy with respect to $\beta$ in the
one-particle orbitals and $A$ in the $1/\sqrt{r}$
part of the two-particle correlation factor which does not
change the phase of the wavefunction. We obtain for
$r_s = 20$ an energy of $-0.0504$ (in $a.u.$) at $\nu = 1/3$ and
$-0.0518$ at $\nu = 1/5$. These are to be compared with
an energy of $-0.0505(1)$ at $\nu = 1/3$ and an
energy of $-0.0518(1)$ at $\nu = 1/5$ from the fixed-phase
diffusion Monte Carlo reported in Ref.~\cite{Ortiz} for
$r_s = 20$ \cite{footnote2}. (See Table IX for our
best energies at these filling factors for $r_s = 20$.)

\section{Energy of a Wigner crystal in the
FQHE regime: Results and Discussions}

In this section,
we present the detailed calculated results and discuss their
possible experimental implications.
There are two experimentally relevant
parameters that determine the energies
of the phases involved here: the filling factor
$\nu$ (related to the carrier density and the magnetic field as
$\nu = 2l_B^2/r_s^2$ when $l_B$ and $r_s$ are in atomic units), and
the electron density parameter $r_s$ (determined by the density,
dielectric screening of the host
media, and the carrier effective mass).
They in turn determine the two relevant energy
scales of the problem: the Landau-level
spacing $\hbar\omega_C$ and the electron-electron interaction
$E_c = e^2 / \varepsilon d$ where $\pi d^2 = 1/n$. Their ratio:
\begin{equation}
{{E_c} \over {\hbar\omega_C}} = \nu r_s / 2,
\label{energyratio}
\end{equation}
provides a measure of the amount of Landau-level-mixing.
For typical 2D electron systems,
$r_s \sim 2$ in atomic units.
But for the $p$-doped systems, it is $r_s \sim 25$, if
$m^* \sim 0.6m_e$ \cite{Stormer83};
and $r_s \sim 13$,
if $m^* \sim 0.3m_e$
\cite{hole}.
For the case of $p$-type doping,
we therefore expect Landau-level-mixing
to play an important role in determining the
energies.

There are two mechanisms by which the WC may lower
its interaction energy by admitting higher Landau-level
components in its wavefunction:
as an inhomogeneous system, both its mean-field Hartree
energy and the dynamic correlation energy can be
lowered.
The former may be done by allowing a charge distribution
more localized than that given by
the lowest Landau-level orbitals, and the latter
by having a nonanalytic correlation
term in the Jastrow factor.
We have therefore
allowed
$\beta$ in Eq.~\ref{wfn} to increase and included the
Landau-level-mixing Jastrow factor in Eq.~\ref{Jastrow1} in
our final variational wavefunction,
in addition to the
magneto-phonon correlations contained in Eq.~\ref{Bij}. Both mechanisms
are found to be important for obtaining an optimal
variational energy.

In Sec.~IVA, we compare the present VMC results with
previous lowest-Landau-level-only calculations.
The energy from a single Slater determinant with
$\beta \ne 1$ is given in Sec.~IVB for $r_s = 20$, and compared
to available Hartree-Fock calculations also with Landau-level-mixing
\cite{LLMHF}. In Sec.~IVC, we show that the
finite size effects are negligible and quantify the
effect of exchange in the present system. In Secs.~IVD, IVE and IVF, we
examine in some detail the
effects of the various variational parameters in our
trial wavefunction on the energy. In order to give a more general
picture for the important effects in a  Wigner crystal, we compare
in Sec.~IVG
the WC energies using several different wavefunctions for $r_s = 20$
and $\nu = 1/3$ and $\nu = 1/5$.
Our calculations are brought into contact with the recent
experiments in Sec.~IVH where we compare the energy of a Wigner
crystal to that of the FQHE liquid and derive a qualitative
phase diagram. Finally, effects of finite temperatures and of
disorder are discussed in Sec.~IVI.

\subsection{Comparison to previous work within the lowest Landau-level
approximation}

We first compare our results within the lowest
Landau-level approximation
to those obtained previously using different methods.
We have constructed the ``Hartree", exchange-only, and
correlation-only
trial wavefunctions and evaluated the respective energies.
The Hartree results
correspond to those using a trial wavefunction that is simply a
product of the one-particle orbitals in Eq.~\ref{wfn} with $\beta = 1$,
with no Jastrow factors.
These can be calculated exactly
by the Ewald summation technique, and this fact is used as an
independent check of the present VMC method.
To include the exchange interaction, we use
a Slater determinant trial wavefunction composed of
the same one-particle orbitals as in the Hartree case.
By ``correlation-only", we mean the wavefunction given by
Eq.~\ref{LGwfn} and Eq.~\ref{Bij}.

The results from these calculations
are given in Table~\ref{Energies}
for $r_s = 2.0$. The
(interaction) energies at other $r_s$'s may
be obtained by simply scaling by
$1/r_s$ since we have implosed the lowest Landau-level approximation.
The size of ``bare" exchange energy may be estimated by
comparing the Hartree and exchange-only results.
In principle, an estimate of the size of exchange, ``screened" by the
magneto-phonon correlations, may be found by
explicitly
antisymmetrizing the wavefunction of Eq.~\ref{LGwfn}.
However, the resultant many-body wavefunction is a sum of
exponentially large number of terms, and cannot be used directly
in an importance sampling calculation.
Fortunately, the upper bound for exchange, set by the ``bare"
exchange interaction, is already very small for $\nu \le {1 \over 3}$.
Landau-level-mixing decreases the exchange
overlap still further, and provides additional screening.
(We shall come back to this point later.)
The kinetic energies for all these wavefunctions
are explicitly evaluated
and confirmed to be exactly ${1\over 2} \hbar\omega_c$ at every step
of the Monte Carlo walks.

We note that our Hartree-Fock (exchange-only) calculation
does not allow as much variational freedom as the previous ones
which consider a charge-density wave
state with the order-parameters $\rho(\vec G)$
being independent at different $\vec G$'s
\cite{WCHF}.
But as can be seen from Table~\ref{Energies},
this difference is not important for total energies.
Our energies from the correlation-only
calculation are also the same as those obtained
by Lam and Girvin \cite{LamGirvin}
using a special $k$-point sampling method. The remaining
differences are within the quoted fitting errors
\cite{LamGirvin}. If we take the
fractional quantum Hall liquid energies from Ref.~\cite{Laughlin,Levesque}
that were obtained also within the lowest Landau-level
approximation, Wigner crystallization
occurs at $\nu_c \sim 1/6.5$
\cite{LamGirvin}.

\subsection{Present results: single Slater
determinant with Landau-level-mixing}

A calculation for the Wigner crystal that is both strictly
variational and includes Landau-level-mixing
is done by MacDonald within the Hartree-Fock
approximation \cite{LLMHF}.
In order to compare with this calculation, we have
calculated the Wigner crystal energy with a single
Slater determinant.
We simply  seek the best one-particle orbital
(optimizing $\beta$ in Eq.~\ref{wfn}) in our VMC calculations.

We give the results for $r_s = 20$
in Table~\ref{HFEnergy} at $\nu = 1/2$, $\nu = 1/3$,
$\nu = 1/5$ and $\nu = 1/7$. The optimal $\beta$ here also
serves as a useful guide for later calculations involving
correlations.
The independent parameters in MacDonald's calculations \cite{LLMHF}
were chosen to be
the filling factor $\nu$ and
the energy ratio $(e^2/\varepsilon l_B) / \hbar\omega_c$ rather than $\nu$ and
$r_s$ as in this work. At $\nu = 1/2$ and $r_s = 20$,
the ratio $(e^2/\varepsilon l_B) / \hbar\omega_c$ is 10, where he also reported
his result. Converting his energy into the present atomic unit,
(at $r_s = 20$ and $\nu = 1/2$,
$e^2/\varepsilon l_B = 0.1~a.u.$, and $\hbar\omega_c = 0.01~a.u.$)
we find that his total energy
(including the ${1 \over 2}\hbar\omega_c$)
is $-0.04311$ (his original result for the total energy was given
without including the
${1 \over 2}\hbar\omega_c$, and as a result is $-0.04811$),
and his total electron-electron
interaction energy is $-0.05022$ \cite{LLMHF}.
{}From the present calculations, we obtain
$-0.0437(1)$ and $-0.0504(1)$, respectively.
There is, of course, no variational principle for the interaction
energy alone, and we do not assign a great significance to its
value; however we can infer from these numbers that the amount of
excess kinetic energy due to Landau-level-mixing from these
two approaches is quite close.
At $r_s = 20$, it is probably insufficient to use only
five lowest Landau-levels as was done in Ref.\cite{LLMHF}.
It appears that
our single Slater determinant wavefunction is quite good
in comparison with the self-consistent Hartree-Fock calculations.

If one ignores exchange, we then have a simple
product of one-particle orbitals. The interaction energy
for this ``Hartree" state can be evaluated rapidly and
accurately with the Ewald summation as we
have done in the lowest Landau-level only cases.
The kinetic energy for this state can also be obtained
straightforwardly by
projecting the one-particle orbitals (for example, take the
one centered at
(0, 0)) onto the $|m\rangle-th$ Landau-level which yields the
following coefficient:
\begin{equation}
c_m = {2\beta \over 1+\beta^2}({1-\beta^2 \over 1+\beta^2})^m.
\label{decomp}
\end{equation}
One can easily verify that
$\sum_0^\infty |c_m|^2 = 1$. The projection also
gives a rough measure of the
size of Landau-level-mixing in cases involving correlations,
although in the latter case projections can not be easily
carried out. For
illustration,
we tabulated the values of $c_m$ for $\beta = 1.3$
in Table~\ref{overlap}.

\subsection{Finite size effects and the size of exchange}

The combination of the Ewald summation technique
and the use of the image charges essentially
eliminated the finite size dependence of the WC energy
using modest size simulation cells.
In Table~\ref{finitesize}, we give the finite size effects
of the present VMC energies.
The results given are for $r_s = 20$, $\nu = 1/3$,
calculated with $\beta = 1.12$ in the one-particle
orbitals, magnotophonon correlation
strenght $A_p = 1.0$ ({\it i.e.}, the original
Lam-Girvin correlation factor), $A = 10.0$ in
the $1 \over \sqrt{r}$ correlation factor, $A/3F^{3/2} = 1/3$,
for
cells of size $6\times 6$, $6 \times 8$,
$8\times 8$, $8\times 10$, $10\times 10$, and $10 \times 12$.

As we mentioned above, a comparison of the Hartree and Hartree-Fock
energies in Table~\ref{Energies} provides a rough estimate of the
size of the bare exchange. It is clear that
for $\nu \le 1/5$, we may safely
ignore the exchange contribution to the total
energy. We now discuss the size of the exchange contribution at
$\nu = 1/3$ near the calculated FQHE liquid-WC transition
in more detail.

Both intra- and inter-Landau-level
correlations can screen the exchange
interaction.
The latter favors smaller
one-particle wavefunctions for a lower direct energy. This reduces
the one-particle wavefunction overlap drastically.
We estimate the
size of the exchange interaction in the WC in the following way.
We form two wavefunctions.
One is
a totally antisymmetric wavefunction formed with
the product of the correlation factor in Eq.~\ref{Jastrow1}
with
a Slater determinant of one-particle orbitals from Eq.~\ref{wfn};
the other one has the same correlation factor, but now has only
one term of the Slater determinant.
We shall refer to the former loosely as
``screened HF",
and the latter
``screened Hartree".
Results are given in
Table~\ref{exchangesize} for $r_s = 20$ and $\nu = 1/3$,
where for comparison the unscreened (``bare") lowest-Landau-level-only
results are listed again.
The ``screened" results are calculated with $\beta = 1.2$ and $A = 10$.
The energies from the ``screened"
wavefunctions are
the same to within the statistical noise. The
magneto-phonon correlation will further screen the exchange interaction.
We thus conclude
that exchange in the Wigner crystal phase
is not important for its energy.

\subsection{
Optimal cusp-condition and the
magneto-phonon correlation}

We have also varied the cusp-conditions given by
$A/3F^{3/2}$ and the strength of the magneto-phonon correlation factor
$A_p$.
The cusp-condition derived from the equation of motion of two particles
in a strong magnetic field is different from that without
the magnetic field. We have considered the short
(relative) distance behavior of two electrons in a strong magnetic
field. We find that for an eigenstate with
relative angular momentum $m$,
the cusp-condition associated with the divergence of the
Coulomb interaction is:
\begin{equation}
{A \over 3F^{3/2}} = {\nu \over {\nu +2}}.
\end{equation}
However, we have checked that changing the cusp-condition at a fixed $A$
from the value of $A/3F^{3/2}=1/3$ by
to up 50 percent at $\nu = 1/3, 1/5$ and $r_s = 20$ does
not affect the energy. This presumably results from the strong
localization of the single-particle orbitals involved.
As a result, the particles are not too close to each
other; and the short range cusp-conditions are not important.
For example, at $\nu =1/5$ and $r_s = 20$, for $\beta = 1.04$,
$A_p=1$, and $A=10$, with $A/3F^{3/2} = 1/3$, the total
energy (with ${1\over 2}\hbar\omega_c$)
is -0.03962(1) $a.u.$, and with $A/3F^{3/2} = 1/5.5$, it is
-0.03961(1) $a.u.$

In Fig.~6, we show the total energy
as a function of the strength of the magneto-phonon
correlation factor $A_p$ for $r_s = 20$ and $\nu = 1/3$.
Here we have set $A = 0$. The
energy changes relatively little
and the optimum occurs for $A_p = 1$. This is not
surprising since $A_p=1$ is the desired value for asymptotically small
$\nu$. Calculations are also done for other filling factors and the
conclusion remains.

\subsection{Effects of $\beta$ on Wigner crystal energy}

We find that for the $r_s$ and $\nu$ values relevant to the
recent experiments carried out on $p$-doped
samples \cite{hole}, squeezing the one-particle
wavefunction is the single most important mechanism for
lowering the energy of a WC \cite{Zhu}. To focus on its effects
on the total energy,
we can set $A_p = 1$ and $A = 0$ and calculate the
total energy as a function of $\beta$.
This was doen in Figs.~1(a) and 1(b) of Ref.~\cite{Zhu}, where we
showed the total energy as a
function of $\beta$ at $\nu = 1/3$ for $r_s = 2$ and
$r_s = 20$ respectively.
When $\beta$ is optimized, the density
at the lattice sites increases by
$\delta \rho(0) / \rho(0) = 70\%$
and the energy is lowered by $\delta E / (E-{1\over 2}\hbar\omega_C)
= -4.4\%$ at $r_s=20$. The changes are
respectively $10\%$ and $ -0.8\%$ at $r_s=2$.
Fig.~7 and Fig.~8 show how the kinetic
energy and the potential energy change with $\beta$
for $r_s=20$ and $\nu = 1/3$.
At the optimal $\beta = 1.3$, the
kinetic energy has risen  only by $\sim 0.001~a.u.$ while
the potential energy gain is $0.0034~a.u.$ compared to
$\beta = 1$. For $\beta$ greater than 1.3, a more rapid rise
in kinetic energy than the drop in potential energy makes it
less favorable, although even at $\beta = 1.6$, the total
energy is still lower than that at $\beta = 1$.
For comparison, we also give the energy at $r_s=20$ and $\nu = 1/5$
as a function of $\beta$ in Fig.~9. The optimal
$\beta$ is around 1.1 and the energy gain is much smaller than that
at $\nu = 1/3$.

We now come back to the case of $r_s = 20$ and $\nu = 1/3$.
A $4\%$ lowering in energy obtained by changing $\beta$ alone
is extremely
important for determining the phase boudary between the FQHE liquid
and the WC.
To see this, we note that the interaction
energy of the lowest Landau-level
Lam-Girvin wavefunction is only 12\% higher than the absolute
minimum set by the Ewald energy.
At $r_s=20$,
this difference is reduced by
one-third, by allowing the Landau-levels to mix
through the one-particle orbitals.
In comparison to the size of the correlation and Landau-level-mixing
effects, the
exchange contributions to the total energies are indeed
negligible.

\subsection{Effects of $A$ and the continuity of $E(\nu)$}

In Fig.~10, we show the total energy of the WC as a function
of $A$, the coefficient of the $1/\sqrt{r}$-term in the Jastrow factor.
The results are calculated for the case of $r_s = 20$, $\nu = 1/3$
with $A_p = 1$ and $\beta = 1.17$.

We have examined the continuity of the WC energy as a function of
filling factor $\nu$ using the VMC method.
We first calculated the WC energies for two filling factors
on both sides of $\nu = 1/3$.
Within the present WC
wavefunction and the numerical accuracy, there is no cusp-like feature
occurring at $\nu = 1/3$.
We also studied the WC energy at
$\nu = 9/20,~1/2,$ and $11/20$.
Again, we find no sign of any discontinuity
at $\nu = 1/2$ within the Wigner crystal wavefunctions.
We therefore conclude that the solid energy curve is continuous
for all filling factors, showing no features at either even- or
odd-denominator filling factors.

\subsection{Energies of a Wigner crystal:
comparison of different wavefunctions }

In Fig.~11
and Fig.~12, we show, for $r_s = 20$, $\nu = 1/3$,
and $r_s = 20$, $\nu = 1/5$,
the VMC energies for a Wigner crystal using
the various different wavefunctions discussed.
These two
figures show
quite clearly which components of the electron correlations
are important and
how they vary with filling factor. For comparison, the
energies of the corresponding Laughlin liquid with and without
Landau-level-mixing from Refs.~\cite{Price93,Levesque} are also given.
For $\nu =1/3$, the largest
energy gain occurs when we allow the Gaussian size to
decrease while keeping the magneto-phonon correlation. Further including
the ${1 \over \sqrt{r}}$ correlation
factor given by Eq.~\ref{Jastrow1} introduces a relatively
small lowering of the energy. We point out that the optimal
single-particle orbital Gaussian
size is larger when the $1/\sqrt{r}$ correlation factor
is included in the wavefunction.
With all the correlation effects taken into account, our
solid energy still lies above that of the liquid (with Landau-level-mixing
effects) at $r_s = 20$. Crystallization does not occur
until a larger $r_s$ at $\nu = 1/3$.

At $\nu = 1/5$, the largest contribution now comes from
intra-Landau-level correlation effects, but still Landau-level-mixing
correlations are comparable in size. Without Landau-level-mixing
effects, the solid energy lies above that of the liquid. But with
Landau-level-mixing effects, the solid becomes lower in energy by
an amount that is significant on the scale of energy differences
in the present context. We therefore conclude
that for $\nu = 1/5$ at $r_s = 20$, the solid is lower in energy than
the FQHE liquid. At even smaller filling factors, the intra-Landau-level
effects are by far the most important effect, and we find that the
Wigner crystal is more stable in energy
(see, however, Ref.~\cite{Platzman93}).

\subsection{Wigner crystal versus FQHE liquid: ground-state energies and
the general phase diagram}

In obtaining our final results for the energy of a Wigner crystal,
we have optimized both $\beta$ and $A$ at a given $r_s$ and $\nu$.
Optimization of $A_p$ is inconsequential for the total
energy. Results for $r_s = 20$ are
plotted in Fig.~13 as solid lines. The energies
from the magneto-phonon correlated wavefunction with no Landau-level-mixing
are shown as empty squares.
The energies from the correlated wavefunction with
Landau-level-mixing are shown as filled squares.
They are obtained by varying $\beta$ in the one-particle orbitals
and $A$ in the $1/\sqrt{r}$ Jastrow factor while keeping
$A_p = 1$ in the magneto-phonon correlation factor.
The energies of the incompressible
FQHE liquid are shown as dotted lines.
Both the lowest-Landau-level results
\cite{Levesque}
based on Laughlin's variational wavefunction,
and the recent results
with Landau-level-mixing
\cite{Price93} are plotted.
For the liquid state, the actual calculated energies at
$\nu = 1/3$, $1/5$, and $1/7$ are plotted
as hexagons in Fig.~13.
The lines passing through them
are a spline fit to the data. They do not show the
cusps that must occur at filling factors where the FQHE states exist.
On the other hand, the energy of the solid is considered to be valid
for all filling factors due to its continuity discussed in Sec.~IVF.

In general, it is expected that Landau-level-mixing
effects will be smaller in the liquid phase.
As discussed in beginning of this section,
in the solid phase, both the Hartree
and correlation energies can be
lowered by allowing Landau-level-mixing.
The former mechanism is found to be
more important for lowering the energy
in WC, but it is entirely lost in the uniform liquid phase
whose Hartree energy will not be altered
by Landau-level-mixing.
This expectation is confirmed by the work of Price, Platzman and He
\cite{Price93}.
They find that
the lowering in the liquid energy is indeed substantially smaller than
that in the solid: for $r_s = 20$ at $\nu = 1/3$, lowering in energy
from Landau-level-mixing for the FQHE liquid state is only  1/4
of that we find for the WC solid (see Fig.~11 and Fig.~13).
As a result, for $r_s=20$, the FQHE state with Landau-level-mixing
is only slightly more stable than the WC state at $\nu =1/3$ for a pure
system with no disorder.
At $\nu = 1/5$, the WC becomes lower in energy.
In Table~\ref{SolidLiquidrs=20}, we list the optimized energies for the Wigner
crystal and the FQHE liquid for $r_s = 20$ at several filling
factors.
For $r_s=2$, the WC state is higher
in energy at $\nu =1/5$ but
remains lower in energy
at $\nu =1/7$ than the FQHE state.

Based on these theoretical results, we present a qualitative
phase diagram for the 2D electron gas in a strong magnetic field.
In Fig.~14, the $x$-axis is the Landau-level filling factor
and the $y$-axis is $r_s$ measured in effective
atomic units which may be changed by
the carrier effective mass at a given doping concentration. At a small
enough filling factor or low
enough density, the system crystallizes.
At intermediate experimental parameters, a
number of reentrant phase transitions at $\nu = 1/5$,
$\nu = 1/3$, $\nu = 1$, {\it etc}, are expected
as one scans the magnetic field. But at a given filling factor,
we only expect one phase transition as $r_s$ is varied.
The picture is only meant to be illustrative; details such as
the strength of various phases (peak heights in the
figure) should not be taken literally.
The details of the phase diagram will be affected by
temperature and disorder, both of which to
lowest order favor the
Wigner crystal phase. We discuss qualitatively
these two issues in the next subsection.

\subsection{Effects of disorder and finite temperatures}

Our calculations are carried out for a perfect 2D electron gas
with no disorder and at zero temperature. Real systems do not
satisfy either condition. Due to the reduced dimentionality,
both are expected to have important effects on the phase diagram
of the system.

The presence of disorder breaks the Wigner crystal into domains
of a finite linear size $\xi^T$ \cite{NLM}. The superscript on
$\xi$ is intended to indicate that most of the static
distortions in a WC are transverse \cite{NLM}.
The Wigner crystal can lower its energy by locally adjusting its
density to accommodate for the local disorder potential. The quantum
Hall liquid on the other hand is rigid to disturbances on energy
scales smaller than the
gap in its collective excitations. Therefore to first order,
the presence of disorder favors the Wigner crystal formation.
The margin by which the Wigner crystal is favored due
to disorder has been estimated \cite{Price93,PriceRev}
using $\xi$ deduced from nonlinear transport threshold field
experiments \cite{I,II,III,Jiang,NLM}. Such a procedure,
while suggestive, is not very accurate.
Quantitatively, the disorder affects
the energy differences between the solid and the liquid,
but the qualitative picture
presented above is found not to be altered \cite{PriceRev}
by such estimates.
The precise pinning mechanism of the Wigner crystal by disorder
in the actual experimental $GaAs/AlGaAS$ heterojunction
systems is at
present not clear, and it remains a subject of considerable amount of
experimental and theoretical interest \cite{Zuzin}.

The effects of finite temperatures on either the Wigner crystal
or the quantum Hall liquid are a difficult issue. On the
fractional quantum Hall
liquid side, while it is often thought that there is only
a gradual decay of the peculiar FQHE order with rising
temperature \cite{FiniteT},
some recent experiments point to the possibility of a finite temperature
transition that is in fact rather abrupt
\cite{GapCollapse}. There is currently no
theoretical model that can account for this observation.
On the Wigner crystal side, a
classical solid in 2D is expected to melt by the Kosterlitz-Thouless
mechanism as the temperature increases \cite{KTmelting}.
A two-step melting path has been
suggested \cite{HalperinNelson}.
The present system has significant quantum
mechanical fluctuation effects and it is unclear if the
classical 2D melting theory is applicable in the vicinity of
a quantum phase transition to the FQHE state.

However, at temperatures well below the afore-mentioned
phase transition
temperatures, the collective excitations in both the
solid and the liquid may be
approximated as independent bosons \cite{Platzman93,GirvinRoton}.
Only the lower branch
needs to be considered, {\it i.e.,} the magneto-roton for
the FQHE liquid and the (largely transverse) magneto-phonon for the
Wigner crystal. In this regime, one can then evaluate, and compare,
the free energies of these two phases, and determine a finite
temperature phase diagram \cite{Platzman93,PriceRev}.
While some interesting effects have
been predicted, it remains unclear if these effects lie outside
the regime of validity of the low-temperature assumption.
The presence of the disorder pinning
gap in the solid phase may also affect the entropy of the
Wigner crystal.
Further theoretical efforts
are needed in these directions.

Finally, we note that
there have been recent claims that Wigner crystallization
occurs for 2D electron gas at doped $Si/SiO_2$ interfaces
at low or zero external magnetic fields \cite{WCSi}.
The reentrant insulating phases set in around $\nu = 1$
(and $\nu = 2$) in these systems. We have carried out calculations
to determine the Wigner crystal/quantum Hall liquid phase
boundary at $\nu = 1$ and found
that the resultant density
is much lower than the experimental doping concentration
reported in these works \cite{WCSi}.
The
band structure of $Si$ around the conduction band minimum is
more complicated than that of $GaAs$ \cite{WCSi}. It is doubtful if
the present model of a one-component fermion system is capable of
describing the $Si/SiO_2$ system.
Theoretically, it has been suggested that multi-component systems are more
susceptible to Wigner crystal or charge-density-wave
instabilities since they may arrange their density profiles
to be mutually beneficial energetically \cite{DoubleLayer},
but this simple picture may not be correct when inter-layer
interaction induced frustration of the triangular lattice
is important.
A variational Monte Carlo calculation has recently been
carried out for the related phenomenon of Wigner crystallization
in a wider quantum well, but within the lowest subband
approximation \cite{FiniteZ}; generalizations of this
approach to double layer systems would be interesting.
While it is possible that such mechanisms are indeed responsible
for the insulating phases observed, the much lower mobility
of the $Si$ samples \cite{WCSi}, as compared to the $GaAs$ samples
\cite{Jiang}, also
brings to mind the possible role of disorder in
the insulating phase.

\section{Summary}

In summary, we have studied variationally the ground-state
energies of two-dimensional electron Wigner crystals, both with and
without an external magnetic field. We identify the important
quantum fluctuations
present in an optimal ground-state wavefunction.
We take into account both the short-range
and long-range correlation effects in the WC
and provide a rigorous upper bound for the WC energy.
Landau-level-mixing
effects are shown to be significant in the range of carrier density,
effective mass, and strength of the
magnetic field of experimental interest.
In the context of the
experimentally observed
insulating phases in the fractional quantum Hall
effect regime, our results strongly suggest that the
main driving force of the phase transitions is electron-electron
interaction in the best current samples.
More work is needed to further understand the effects of finite
temperatures and of disorder.
This will probably require a better understanding of both
the neutral and charged excitations of this interesting
electron solid.

\acknowledgments

We thank Drs. Sankar Das Sarma, Song He,
Pui Lam, Peter Littlewood, Andy Millis,
Phil Platzman, Rod Price, and Andrew Rappe
for discussions and collaborations on related subjects.
Work at Berkeley was supported
by NSF through Grant No. DMR91-20269 and by
DOE through Contract No. DE-AC03-76SF00098.
CRAY computer time was provided
by the NSF at the San Diego Supercomputing
Center and by the DOE.
Work at Rutgers was supported by NSF through
DMR92-21907.

\newpage
\begin{table}
\caption
{Energy/electron (in $10^{-3}$ atomic units) of the Wigner crystal
in a 2D hexagonal lattice from the
present VMC calculation in comparison with previous work.
Results are for a simulation cell of 56 electrons.
}
\begin{tabular}{lccc}
           & $r_s = 30$ & $r_s = 50$ & $r_s=100$ \\
\tableline
 VMC-Present & -31.83(1) & -19.82(1) & -10.23(1) \\
 VMC$^a$     & -31.82(1) &-19.79(1) & -10.24(1) \\
             &           &          &           \\
 GFMC$^a$    &-31.89(1)  &-19.81(1) &-10.24(1)   \\
\end{tabular}
$^a$ Ref.~\cite{Ceperley}
\label{EnergiesB=0}
\end{table}

\begin{table}
\caption
{Energy/electron (in $10^{-3}$ atomic units) for electrons
in hexagonal (HX), square (SQ), and honeycomb (HC) lattices.
Results are for the spin-polarized state.
Sizes of the simulation cells are slightly
different for each lattice and in every case the
resulting finite size effects are smaller than the
statistical noise (see Table~III).
}
\begin{tabular}{ccccc}
    & $r_s = 30$ &$r_s = 50$ &$r_s = 70$ & $r_s =100$ \\
\tableline
HX & -31.83(1) & -19.82(1) &-14.38(1) & -10.23(1) \\
SQ & -31.72(1) & -19.71(1) & -14.34(1) & -10.19(1) \\
HC & -31.52(1) &-19.60(1) & -14.25(1) & -10.13(1) \\
\end{tabular}
\label{VMCEnergies}
\end{table}

\begin{table}
\caption
{Finite size effects (energy/electron
in $10^{-3}$ atomic units)
for electrons in the
honeycomb (HC) lattice  and the square (SQ) lattice at
$r_s=30$.}
\begin{tabular}{cccc}
         & N=32    & N=50    & N=72 \\
\tableline
\multicolumn{4}{c}{honeycomb}\\
\\
FM & -31.50(1) & -31.51(1) &-31.51(1)\\
AFM& -31.67(1) & -31.67(1) &-31.67(1) \\
\tableline
\\
\tableline
\tableline
         &N=36 &N=48 &N=64 \\
\tableline
\multicolumn{4}{c}{square} \\
\\
FM &-31.71(1) &-31.72(1) &-31.72(1) \\
AFM &-31.67(1) &-31.67(1) &-31.67(1) \\
\end{tabular}
\label{finitesizeB=0}
\end{table}

\begin{table}
\caption
{Energies (in effective atomic units)
of the hexagonal Wigner crystal from various lowest
Landau-level only calculations at $r_s =2.0$.
A constant kinetic energy
${1\over 2}\hbar \omega_C$ is subtracted.
}
\begin{tabular}{ccccccc}

$\nu$ &Hartree & \multicolumn{2}{c} {Exchange-only}
               & \multicolumn{2}{c} {Correlation-only}             &Laughlin \\
      &        &Present
               &Ref.~\cite{WCHF,LamGirvin} &Present
               &Ref.~\cite{LamGirvin} &Ref.~\cite{Levesque}   \\
\tableline
 1/2  &-0.4222 & -0.4435(3) & -0.4438 & -0.4397(13) & -0.4396 &\\
 1/3 &-0.4722 & -0.4762(8) & -0.4758 & -0.4834(9)  &-0.4836 &-0.5023 \\
 1/4 &-0.4960 & -0.4966(5) & -0.4957 & -0.5040(3)  &-0.5034 &\\
 1/5 &-0.5090 & -0.5091(1) & -0.5090 & -0.5155(3)  &-0.5151 &-0.5180\\
 1/7 &-0.5226 & -0.5225(1) & -0.5220 & -0.5272(1)  &-0.5264 &-0.5256\\
\end{tabular}
\label{Energies}
\end{table}

\begin{table}
\caption
{Wigner crystal energy (in atomic units per
electron) for a single Slater determinant
with one-particle orbitals given in Eq.~18, for
$r_s = 20$.
The optimal variational parameter $\beta$
at which these calculations are done is also given.
}
\begin{tabular}{cccc}
Filling factor $\nu$ & $\beta$  & Total Energy &${1 \over 2}\hbar\omega_c$\\
\tableline
1/2 & 1.5  & -0.0437(1) & 0.0050 \\
1/3 & 1.37  & -0.0424(1) & 0.0075 \\
1/5 & 1.15  & -0.0391(1) & 0.0125 \\
1/7 & 1.075 & -0.0350(1) & 0.0175 \\
\end{tabular}
\label{HFEnergy}
\end{table}

\begin{table}
\caption
{Overlap of $\phi(\beta)$ in Eq.~18 with the
$nth$-LL: $<nth~LL|\phi(\beta)>$ with $\beta=1.3$.
This gives a measure of the amount of Landau-level-mixing in the
ground-state wavefunction corresponding to a Wigner crystal at
$r_s = 20$ and $\nu = 1/3$ (see text).}
\begin{tabular}{cccc}
$n = 0$ & $n = 1$ & $n = 2$ & $n=3$ \\
\tableline
0.9666 & -0.2480 & 0.0636 & -0.0163 \\
\end{tabular}
\label{overlap}
\end{table}

\begin{table}
\caption
{Finite size effects and statistical
noise in this work have been reduced to a level that is
unimportant for the energy differences of
interest. We show a typical example for $\nu = 1/3$,
$r_s = 20$, calculated with $A = 10.0$, $A_p = 1.0$, $A/(3F^{3/2}) = 1/3$,
$\beta = 1.12$. See text for these parameters in
the wavefunction.
The total energies here are in atomic units
and include the term ${1 \over 2}\hbar\omega_c$ = 0.0075 $a.u.$}
\begin{tabular}{cccc}
Simulation cell size & Energy/electron  \\
\tableline
6  $\times$ 6  & -0.04322(1) \\
8  $\times$ 6  & -0.04322(1) \\
8  $\times$ 8  & -0.04322(1) \\
10 $\times$ 8  & -0.04322(1) \\
10 $\times$ 10 & -0.04322(1) \\
12 $\times$ 10 & -0.04321(1) \\
\end{tabular}
\label{finitesize}
\end{table}

\begin{table}
\caption
{An estimate for the size of exchange
contribution to the total energy: comparison of screened Hartree
with screened Hartree-Fock (HF)
for $r_s = 20$ at
$\nu = 1/3$.
For comparison, the unscreened results from
Table~IV are also shown here.
Total energies (not including ${1 \over 2} \hbar\omega_c$)
are given in atomic units.
}
\begin{tabular}{ccccc}
Bare Hartree & Bare HF & Screened Hartree & Screened HF  \\
\tableline
 -0.04722 & -0.04762(9) & -0.05035(1) & -0.05035(1) \\
\end{tabular}
\label{exchangesize}
\end{table}

\begin{table}
\caption
{Ground state energies at $r_s = 20$ for
the Wigner solid versus the FQHE liquid from Ref.~[7]
in atomic units. We have taken out
the term ${1 \over 2}\hbar\omega_c$. For its value at
respective filling factors, see Table~V.
}
\begin{tabular}{cccc}
Filling factor $\nu $ & Wigner crystal & FQHE liquid   \\
\tableline
1/3 & -0.05073(1) & -0.05090(1) \\
1/5 & -0.05212(1) & -0.05203(1) \\
1/7 & -0.05291(1) & -0.05265(1) \\
1/9 & -0.05341(1) & -0.05306(1) \\
\end{tabular}
\label{SolidLiquidrs=20}
\end{table}

\newpage

\figure{Fig.~1~$E_{FM}-E_{AFM}$
(in $10^{-3}$ atomic units)
for electrons in
a square lattice. The line
is a spline fit to guide the eye. The FM state has a lower energy.
\label{fig.1} }

\figure{Fig.~2~$E_{FM}-E_{AFM}$ (in $10^{-3}$ atomic units)
for electrons in
a honeycomb lattice. The line
is a spline fit to guide the eye. The AFM state has a lower energy.
\label{fig.2} }

\figure{Fig.~3~$E_{FM}-E_{AFM}$
(in $10^{-3}$ atomic units)
as a function of $r_G$, the width of the
Gaussians in the one-particle orbitals, for electrons in the
honeycomb lattice at $r_s=30$. The parameters A and F
in the Jastrow factor
are not altered with $r_G$.
\label{fig.3} }

\figure{Fig.~4~$E_{FM} - E_{AFM}$
(in $10^{-3}$ atomic units)
as a function of aspect ratio
$a/b$ for the rectangular lattice at $r_s=30$.
Gaussians in the
one-particle orbitals are anisotropic
(see text for details).
\label{fig.4} }

\figure{Fig.~5~Schematic illustration of Thouless
and Li's \cite{RingExThouless}
argument for the sign of the ring-exchanges
in a strong magnetic field. The underlying assumption
is that the total energy increases with the number of nodes
in the spatial part of the wavefunction. It is argued, in the text,
that this assumption does not hold in the FQHE regime.
\label{fig.5} }

\figure{Fig.~6~ Total energy (in effective atomic unit)
as a function of $A_p$, the strength
of the magneto-phonon correlation, for
$\nu = 1/3$ at $r_s = 20$. The line is a spline fit.
A constant $1 \over 2$$\hbar\omega_c$ is subtracted.
\label{fig.6} }

\figure{Fig.~7~ Kinetic energy (in atomic units) per electron
vs $\beta$ for $\nu = 1/3$ at $r_s = 20$
in the effective atomic unit.  The line is a spline fit.
\label{fig.7} }

\figure{Fig.~8~ Interaction energy (in atomic units) per electron
vs $\beta$ for $\nu = 1/3$ at $r_s = 20$
in the effective atomic unit.  The line is a spline fit.
\label{fig.8} }

\figure{Fig.~9~ $E^{total} - {1\over 2}\hbar\omega_C$ (per electron)
vs $\beta$ for $\nu = 1/5$ at $r_s = 20$.
Energies are
in the effective atomic unit.
Statistical noise are given by the size of the data points.
The line is a spline fit.
\label{fig.9} }

\figure{Fig.~10~
$E^{total} - {1\over 2}\hbar\omega_C$ (per electron) vs $A$
for $\nu = 1/3$ and $r_s = 20$ calculated with $A_p = 1$
and $\beta = 1.17$.
Energies are
in the effective atomic unit.
Statistical noise are given by the size of the data points.
The line is a spline fit.
\label{fig.10} }

\figure{Fig.~11~ WC solid and the FQHE liquid energies
at $\nu = 1/3$ and $r_s = 20$
in the effective atomic unit
calculated from different wavefunctions. (The
liquid remains the ground-state at $r_s = 20$.)
WC1: Hartree-Fock results with no Landau-level-mixing.
WC2: Magneto-phonon correlated results with no Landau-level-mixing.
WC3: Lowest possible energy without introducing the Landau-level-mixing
Jastrow factor, but allows Landau-level-mixing by changing $\beta$.
WC4: Lowest energy for the WC with all variational parameters
optimized.
FQHE-1: FQHE liquid energy from the Laughlin wavefunction with no
Landau-level-mixing \cite{Levesque}.
FQHE-2: FQHE liquid energy with
Landau-level-mixing from Ref.~\cite{Price93}.
\label{fig.11} }

\figure{Fig.~12~ FQHE liquid and WC solid energies
at $\nu = 1/5$ at $r_s = 20$
in the effective atomic unit
calculated from different wavefunctions. (The
solid has a lower energy at $r_s = 20$.)
The notations for the data points are identical to those in Fig.~11.
\label{fig.12} }

\figure{Fig.~13~$E^{total} - {1\over 2}\hbar\omega_C$
of the WC at $r_s = 20$ compared with those of the FQHE liquid.
Energies are in effective atomic unit.
Heavy line connecting the empty squares is the WC energy
with no Landau-level-mixing but with magneto-phonon correlations.
Heavy line connecting the filled squares is the WC energy
with all correlation effects considered obtained in this work.
Dotted line connecting the empty hexagons is
the Laughlin state with no Landau-level-mixing from Ref.~[10].
Dotted line connecting the filled hexagons is
the Laughlin state with Landau-level-mixing from Ref.~[16].
\label{fig.13} }

\figure{Fig.~14~ Schematic phase diagram of a 2D electron/hole
system at $T = 0$ and free of disorder. The $x$-axis
is filling factor and the $y$-axis is the effective
$r_s$. Several possible reentrant phase trsnsitions around the
principles FQHE states are illustrated. We have
assumed that at $\nu = 1/7$ the ground-state is the Wigner crystal.
\label{fig.14} }


\begin{references}
\bibitem{footnote1}
In order to do so, the electrons must now be treated as
distinguishable particles as we discussed in the
previous subsection, Sec.~IIID. This however
has little effect on the WC total energy, see Sec.~IVC below for a
detailed quantitative exposition.
\bibitem{footnote2}
The uncertainties in the fixed-phase results are from the
fact that they are read off from Fig.~3 in Ref.~\cite{Ortiz}.
The energies here do not include the term of
${1 \over 2}\hbar\omega_c$.
\end{references}
\end{document}